\documentclass{article}
\usepackage{amssymb}
\usepackage{amsbsy}
\usepackage{rotating}
\newcommand{\be}{\begin{equation}}
\newcommand{\ee}{\end{equation}}
\newcommand{\Tr}{{\rm Tr}}
\def\bea{\begin{eqnarray}}
\def\eea{\end{eqnarray}}
\def\bean{\begin{eqnarray*}}
\def\eean{\end{eqnarray*}}

\newcommand{\barr}{\begin{array}}
\newcommand{\earr}{\end{array}}

\newcommand{\bed}{\begin{displaymath}}
\newcommand{\eed}{\end{displaymath}}
\newcommand{\bal}{\begin{array}{ll}}
\newcommand{\eal}{\end{array}}

\def\bvec#1{\raise1.5ex\hbox{$\rightarrow$}\mkern-16.5mu #1}

\def\ket#1{\vert\,#1>}
\def\m#1{\mathcal#1}
\newcommand{\bs}{\boldsymbol}

\begin{document}

\title{\hfill ~\\[-30mm]
       \hfill\mbox{\small UFIFT-HEP-07-13}\\[30mm]
       \textbf{Simple Finite Non-Abelian \\Flavor Groups }}
\date{}
\author{\\Christoph Luhn${}^1$,\footnote{E-mail: {\tt luhn@phys.ufl.edu}}~~
        Salah Nasri${}^{1,2}$,\footnote{E-mail: {\tt snasri@phys.ufl.edu}}~~
        Pierre Ramond${}^1\,$\footnote{E-mail: {\tt ramond@phys.ufl.edu}}\\ \\
\emph{\small{${}^1$}Institute for Fundamental Theory, Department of Physics,}\\
\emph{\small University of Florida, Gainesville, FL 32611, USA}\\[4mm]
\emph{\small{${}^2$}Department of Physics, United Arab Emirates University,}\\
\emph{\small P.O.Box 17551, Al Ain, United Arab Emirates}\\[3mm]}

\maketitle

\begin{abstract}
\noindent The recently measured unexpected neutrino mixing patterns have caused a resurgence of interest in the study of finite flavor groups with two- and three-dimensional irreducible representations. This paper details the mathematics of the two finite {\em simple} groups with such representations,  the Icosahedral group $\m A_5$, a subgroup of $SO(3)$, and $\mathcal{PSL}_2(7)$, a subgroup of $SU(3)$. \end{abstract}

\thispagestyle{empty}
\vfill
\newpage
\setcounter{page}{1}


\section{Introduction}
The gauge interactions of the Standard Model are associated with symmetries which naturally generalize to  a Grand-Unified structure. Yet there is no corresponding recognizable symmetry in the Yukawa couplings of the three chiral families.  I.~I.~Rabi's old question about the muon,  ``Who ordered it?", remains unanswered. Today, in spite of the large number of measured masses and mixing angles, the origin of the chirality-breaking Yukawa interactions remain shrouded in mystery. 

A natural suggestion that the three chiral families assemble in three- or
two-dimensional representations of a continuous group, single out  $SU(3)$
\cite{su3}, or $SU(2)$ \cite{su2}, respectively, as natural flavor groups,
which have to be broken at high energies to avoid flavor-changing neutral
processes. Unfortunately, such hypotheses did  not prove particularly
fruitful.  

However, the recently measured MNSP lepton-mixing matrix displays a strange
pattern with two large and one small rotation angle. The near zero in one of
its entries suggests a non-Abelian finite symmetry [3\,-\,8].
Indeed a matrix which approximates the
measured mixing appears naturally in the smallest non-Abelian finite group,
$\m S_3$. 

Since there are three chiral families, it is natural to concentrate on those finite group which have two and three dimensional irreducible representations:   the possible finite flavor groups must be finite subgroups of continuous $SU(2)$, $SO(3)$, and $SU(3)$. These finite subgroups were all identified early on by mathematicians, and are listed in a ninety years old textbook \cite{oldbook}.

One finds only two {\em simple} groups in their list. One is the Icosahedral
group, a subgroup of $SO(3)$; the other is $\mathcal{PSL}_2(7)$, a subgroup of
$SU(3)$. There are no finite {\em simple} subgroups of $SU(2)$.  The purpose
of this paper is to study in some detail these groups, in particular
$\mathcal{PSL}_2(7)$. For their application in quiver theories, see
\cite{Hanany:1998sd}. 

\section{Finite Groups with Two-dimensional Irreps}
The finite groups with two-dimensional representations  were identified as early as 1876 by Felix Klein, who considered the fractional linear transformations on two complex variables 
 
\be Z'~=~\pmatrix{z_1'\cr z_2'}~=~\m M\,Z~=~\pmatrix{a&b\cr -\overline b& \overline a}\pmatrix{z_1\cr z_2}\ ,\ee
where $\m M$ is a unitary matrix with unit determinant. He then constructed the three quadratic forms, shown here in modern notation, (vector analysis had not yet been invented) 
 
\be\vec r~=~Z^\dagger\vec \sigma\,Z\ ,\ee
where $\vec\sigma$ are known today as the Pauli (not Klein) matrices. He clearly understood how to go from $SU(2)$ to $SO(3)$, as he realized that the transformations acted on the vector $\vec r$ as rotations about the origin. These are five types of rotations, 

\begin{itemize}

\item rotations of order $n$ about one axis; they  generate the finite Abelian group $\m Z_n$. 
 
 \item rotations with two rotation axes which generate the Dihedral group $\m D_{n}$ with $2n$ elements.
 
 \item rotations which map the tetrahedron into itself form $\m T=\m A_4$, the Tetrahedral group  with $12$ elements.
 
 \item rotations which map the octahedron into itself form $\m O=\m S_4$, the Octahedral group with  $24$ elements.
 
 \item rotations which map the icosahedron (dodecahedron) into itself form $\m I=\m A_5$, the Icosahedral group of order $60$, the only {\em simple} finite rotation group.
 
 \end{itemize}
    
\noindent The Dihedral groups, the symmetry groups of plane polygons, have
only singlet and {\em real} doublet representations ($n>2$). The last three,
$\m T$, $\m O$ and $\m I$,  are symmetry groups of the Platonic solids. As
$SO(3)$ subgroups, all three have real triplet representations, but no
two-dimensional spinor-like representations. Each can be thought of as the
quotient group of  its  {\em binary}-equivalent group with at least one
two-dimensional ``spinor" representation and twice the number of elements
(double cover). Below we give in some detail the construction of the
Icosahedral group and of its binary form.     

\vskip .3cm
\noindent{\bf Icosahedral Group $\bs{\m A_5}$}
\vskip .3cm

\noindent We begin with a description of $\m A_5=\m I$, the only simple finite $SO(3)$ subgroup. Its sixty group elements are generated by two elements, $A$ and $B$,  with two equivalent presentations
\bea
&&<\!A, B\,|\,A^2=B^3=(AB)^5=1\!>\ ,\nonumber \\[1mm]
&& <\!A, B, C\,|\,A^2=B^3=C^5=(ABC)=1\!> .\eea
It is isomorphic to $\mathcal{PSL}_2(5)$, the group of projectively defined $(2\times 2)$ matrices of unit determinant over $\mathbb F_5$, the finite Galois field with five elements. It is the symmetry group of the icosahedron, with $12$ rotations of order $5$ (by $72^\circ$), $12$ rotations of order $5$ (by $144^\circ$), $20$ rotations of order $3$ (by $60^\circ$), and $15$ rotations of order $2$ (by $180^\circ$). These correspond to its conjugacy classes, displayed in its character table (the prefactor and the square bracket denote the number and the order\footnote{The order of an element of a finite
group is the smallest power needed to obtain the identity element.} of the elements in the class, respectively),

\vskip .3cm
{\footnotesize{\begin{center}
\begin{tabular}{c|ccccc}
&   \\
{\large$\bs {\m A_5}$}~~& $C_1$&\hfill $12C^{[5]}_2$& \hfill$12C^{[5]}_3$&\hfill $15C^{[2]}_4$&\hfill $20C^{[3]}_5$\\
 &&&&&   \\
\hline    
 &&&&&   \\
$\chi^{[\bf 1]}_{}$~~& $1$&\hfill $1$&\hfill  $1$&\hfill  $1$&\hfill $1$\\
 &&&&&   \\
$\chi^{[\bf 3_1]}_{}$~~& $3$&\hfill  $b^{}_5$&\hfill  $b_5^{'}$&\hfill  $-1$&\hfill $0$ \\
 &&&&&   \\
$\chi^{[\bf 3_2]}_{}$~~& $3$&\hfill  $b_5^{'}$&\hfill  $b_5^{}$&\hfill  $-1$&\hfill $0$\\
 &&&&&   \\
$\chi^{[\bf 4]}_{}$~~& $4$&\hfill  $-1$&\hfill  $-1$&\hfill  $0$&\hfill $1$\\
 &&&&&   \\
$\chi^{[\bf 5]}_{}$~~& $5$& \hfill $0$&\hfill  $0$&\hfill  $1$&\hfill $-1$\\
\end{tabular}\end{center}
}}
\vskip .5cm

\noindent with 
\be b_5~=~\frac{1}{2}(1+\sqrt{5})\ ,\qquad b_5'~=~\frac{1}{2}(1-\sqrt{5})\ .\ee
It has five real irreducible representations, two of which are triplets. We compute their Kronecker products, using the following general method: given two irreps $\bf r$ and $\bf s$, with Kronecker product 
\be
{\bf r} \otimes {\bf s} ~=~ \sum_{\bf t} d({\bf r},{\bf s},{\bf t}) \, {\bf t}
\ ,
\ee
the positive integers $d({\bf r},{\bf s},{\bf t})$ are given by the formula
\be
d({\bf r},{\bf s},{\bf t}) ~=~ \frac{1}{N} \, \sum_{i} n^{}_i \cdot
\chi_i^{[{\bf r}]} \, \chi_i^{[{\bf s}]} \, \overline{\chi}_i^{[{\bf t}]} \ .
\ee
Here, $N$ is the order of the group, the sum is over all classes and $n_i$ denotes the corresponding
number of elements in the class.  Applied to $\m A_5$, we obtain the following table, where we indicate the 
symmetric and the antisymmetric products of ${\bf r}\otimes {\bf r}$ by the subscripts $s$ and $a$, respectively.

\vskip .3cm
\begin{center}
{{\begin{tabular}{|c|}
 \hline  \\
~{\bf $\bs {\m A^{}_5}~$  Kronecker Products}\hfill\\
   \\
\hline    
 \\
${\bf 3_1} \,\otimes\, {\bf 3_1}=~{\bf 3_1}_a~+~({\bf 1}~+~{\bf 5})_s\hfill$ \\
${\bf 3_2}\,\otimes\, {\bf 3_2}=~{\bf 3_2}_a~+~({\bf 1}~+~{\bf 5})_s\hfill$ \\
$ {\bf 3_1}\,\otimes\, {\bf 3_2} =~{\bf 4}~+~{\bf 5}\hfill$ \\ 
${\bf 3_1}\,\otimes\, {\bf 4}\,~=~{\bf 3_2}~+~{\bf 4}~+~{\bf 5}\hfill$ \\
${\bf 3_2}\,\otimes\, {\bf 4}\,~=~{\bf 3_1}~+~{\bf 4}~+~{\bf 5}\hfill$ \\
${\bf 3_1 }\,\otimes\, {\bf 5}\,~=~{\bf 3_1}~+~{\bf 3_2}~+~{\bf 4}~+~{\bf 5}\hfill$ \\
${\bf 3_2}\,\otimes\, {\bf 5}\,~=~{\bf 3_1}~+~{\bf 3_2}~+~{\bf 4}~+~{\bf 5}\hfill$ \\
$~\,{\bf 4}\,\otimes\, {\bf 4}\,~=~({\bf 3_1}~ +~{\bf 3_2})_a~+~({\bf 1}~+~{\bf 4}~ +~{\bf 5})_s\hfill$ \\
$~\,{\bf 5}\,\otimes\, {\bf 5}\,~=~({\bf 3_1}~ +~{\bf 3_2}~ +~{\bf 4})_a~ +~({\bf 1}~+~{\bf 4}~ +~{\bf 5}~ +~{\bf 5})_s\hfill$ \\
$~\,{\bf 4}\,\otimes\, {\bf 5}\,~=~{\bf  3_1}~ +~{\bf 3_2}~ +~{\bf 4}~ +~{\bf 5}~ +~{\bf 5}\hfill$ \\ \\ \hline
\end{tabular}}}\end{center}
\vskip .5cm
\noindent{\bf Binary Icosahedral Group $\bs{\mathcal{SL}_2(5)}$}
\vskip .3cm
\noindent  Starting from $\m A_5=\mathcal{PSL}_2(5)$, we can construct a
closely related group by dropping  the projective restriction on the
matrices. This yields $\mathcal{SL}_2(5)$, a group with twice as many
elements, the binary form of $\m A_5$ (see e.g. \cite{patera}). This group is no longer simple since it has a normal subgroup of order two. 

Its presentation is of the same form as for $\m A_5$ 
\be
<\,A, B, C\,|\,A^2=B^3=C^5=(ABC)\,>\ ,\ee
but here the element $A^2$ is {\bf not} the identity, but in some sense the negative of the identity, so that its square is the identity element. Hence it has only one element of order two, and $\m A_5$ is not a subgroup, but its largest {\em quotient} group.   

In terms  of representations, it corresponds to adding  (at least) one two-dimensional {\em spinor} representation to those of $\m A_5$.  In that representation, the generators are given by
\be
C=\pmatrix{-\chi^2&0\cr 0&-\chi^3}\ ,~\quad
B=\frac{1}{\sqrt{5}}\pmatrix{\chi^4-\chi^2&\chi-1\cr 1-\chi^4&\chi-\chi^3}\
,~\quad A=B\,C \ ,\ee
where
\be
\chi^5=1\ .\ee
$ \mathcal{SL}_2(5)$ has in fact  four spinor-like representations, $\bf 2_s$, $\bf 2_s'$, $\bf 4_s$, and $\bf 6_s$.  The character table is given by

\vskip .3cm
{\footnotesize{\begin{center}
\begin{tabular}{c|ccccccccc}
&   \\
{\large$\bs {\mathcal{SL}_2(5)}$}~~& $C_1$&\hfill $1C^{[2]}_2$&\hfill $12C^{[5]}_3$& \hfill$12C^{[5]}_4$&\hfill $12C^{[10]}_5$&\hfill $12C^{[10]}_6$&\hfill $20C^{[3]}_7$&\hfill $20C^{[6]}_8$&\hfill $30C^{[4]}_9$\\
 &&&&&&&&&   \\
\hline    
 &&&&&&&&&   \\
$\chi^{[\bf 1]}_{}$~~& $1$&\hfill $1$&\hfill  $1$&\hfill  $1$&\hfill $1$&\hfill $1$&\hfill $1$&\hfill $1$&\hfill $1$\\
 &&&&&   \\
$\chi^{[\bf 3_1]}_{}$~~& $3$&\hfill  $3$&\hfill  $b^{}_5$&\hfill  $b^{'}_5$&\hfill $b_5'$&\hfill $b_5^{}$&\hfill $0$&\hfill $0$&\hfill $-1$ \\
 &&&&&   \\
$\chi^{[\bf 3_2]}_{}$~~& $3$&\hfill  $3$&\hfill  $b^{'}_5$&\hfill  $b^{}_5$&\hfill $b^{}_5$&\hfill $b_5'$&\hfill $0$&\hfill $0$&\hfill $-1$\\
 &&&&&   \\
$\chi^{[\bf 4]}_{}$~~& $4$&\hfill  $4$&\hfill  $-1$&\hfill  $-1$&\hfill $-1$&\hfill $-1 $&\hfill $1$&\hfill $1$&\hfill $0$\\
 &&&&&   \\
$\chi^{[\bf 5]}_{}$~~& $5$&\hfill  $5$& \hfill $0$&\hfill  $0$&\hfill $0$&\hfill $0$&\hfill $-1$&\hfill $-1$&\hfill $1$\\
 &&&&&   \\
 $\chi^{[\bf 2_s]}_{}$~~& $2$&\hfill  $-2$& \hfill $-b_5^{}$&\hfill  $-b_5'$&\hfill $b_5'$&\hfill $b_5^{}$&\hfill $-1$&\hfill $1$&\hfill $0$\\
 &&&&&   \\
 $\chi^{[\bf 2_s']}_{}$~~& $2$&\hfill  $-2$& \hfill $-b_5'$&\hfill  $-b_5^{}$&\hfill $b_5^{}$&\hfill $b_5'$&\hfill $-1$&\hfill $1$&\hfill $0$\\
 &&&&&   \\
 $\chi^{[\bf 4_s]}_{}$~~& $4$&\hfill  $-4$& \hfill $-1$&\hfill  $-1$&\hfill $1$&\hfill $1$&\hfill $1$&\hfill $-1$&\hfill $0$\\
 &&&&&   \\
 $\chi^{[\bf 6_s]}_{}$~~& $6$&\hfill  $-6$& \hfill $1$&\hfill  $1$&\hfill $-1$&\hfill $-1$&\hfill $0$&\hfill $0$&\hfill $0$\\
\end{tabular}\end{center}
}}
\vskip .5cm

\newpage
\noindent We can easily find the Kronecker products of its irreps, by using
the usual technique, but for the finite subgroups of $SU(2)$, there is another
method which relies on the {\em McKay Correspondence} \cite{McKay}. This
theorem states that Kronecker products are related to extended Dynkin
diagrams. In the following we illustrate this relation for $\m S \m L_2(5)$.

Any finite subgroup of $SU(2)$ will have (at least) one spinor doublet $\bf 2_s$, and of course one singlet irrep.  We are going to generate its nine irreducible representations by taking repeated products with the spinor doublets, $\bf 2_s$. Start with the singlet  

$$
\bf 1\otimes 2_s~=~2_s\ ,$$
which we represent graphically as follows: assign a dot to the $\bf 1$ and one dot the the $\bf 2_s$ which appears on the right-hand-side. We connect these two dots with a line, to indicate that they are obtained form one another by taking the product with $\bf 2_s$.  We repeat the process  with the right-most dot

$$
\bf 2_s\otimes \bf 2_s~=~\bf 1~\oplus~\bf 3_1\ ,$$
which yields a third dot for the $\bf 3_1$, connected to the $\bf 2_s$ dot by a line. So far we have a linear diagram of three dots connected by a line. Next we consider the product $\bf 2_s\otimes \bf 3_1$, which has to be a sum of  spinorial representations. They are uniquely determined by  the McKay Correspondence, which says that the diagram obtained in this way is   the $E_8$ extended Dynkin diagram! 

\vskip .5cm
\setlength{\unitlength}{.013in}
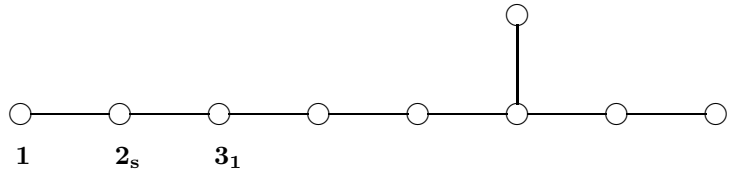
\begin{picture}(300,100)
\put(14,50){\line(1,0){32}}
\put(54,50){\line(1,0){32}}
\put(94,50){\line(1,0){32}}
\put(134,50){\line(1,0){32}}
\put(174,50){\line(1,0){32}}
\put(214,50){\line(1,0){32}}
\put(254,50){\line(1,0){32}}
\put(210,54){\line(0,1){32}}
\put(10,50){\circle{8}}
\put(50,50){\circle{8}}
\put(90,50){\circle{8}}
\put(130,50){\circle{8}}
\put(170,50){\circle{8}}
\put(210,50){\circle{8}}
\put(250,50){\circle{8}}
\put(210,90){\circle{8}}
\put(290,50){\circle{8}}
\put(8,30){\shortstack{$\bf 1$}}
\put(48,30){\shortstack{$\bf 2_s$}}
\put(88,30){\shortstack{$\bf 3_1$}}
\end{picture}
\vskip .3cm
\noindent Hence there must be a $\bf 4_s$ at the fourth dot, that is 

$$
\bf 2_s\otimes \bf 3_1~=~ \bf 2_s+\bf 4_s\ . $$
Similarly,  

$$
\bf 2_s\otimes \bf 4_s~=~\bf 3_1+\bf 5\ ,$$
yields a sum of tensor irreps which continues the linear chain. The product

$$
\bf 2_s\otimes \bf 5~=~\bf 4_s+\bf 6_s\ ,$$
adds one more rung to the linear chain. Next, since the right-hand side can only contain tensor irreps, with a unique solution 

$$
\bf 2_s\otimes\bf 6_s~=~\bf 5+\bf 3_2+\bf 4\ ,$$
 we see that the linear chain breaks into two branches. The dot on the first branch  is $\bf 3_2$, and it terminates, indicating that

$$
\bf 2_s\otimes \bf 3_2~=~\bf 6_s\ .$$ The second branch keeps going,
requiring that

$$
\bf 2_s\otimes \bf 4~=~\bf 6_s+\bf 2_s'\ .$$
 The last dot associated with $\bf 2_s'$ terminates in order to satisfy the McKay Correspondence, that is

$$
\bf 2_s\otimes \bf 2_s'~=~\bf 4\ .$$
 This completes the process, with one irrep associated with each dot: 
 
\vskip .5cm
\setlength{\unitlength}{.013in}
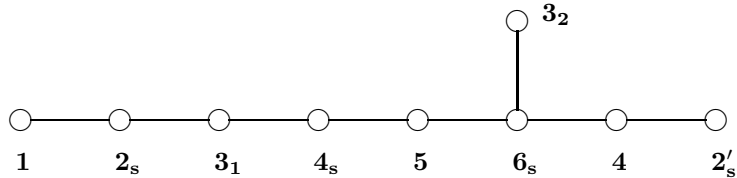
\begin{picture}(300,100)
\put(220,90){\shortstack{$ \bf 3_2$}}
\put(14,50){\line(1,0){32}}
\put(54,50){\line(1,0){32}}
\put(94,50){\line(1,0){32}}
\put(134,50){\line(1,0){32}}
\put(174,50){\line(1,0){32}}
\put(214,50){\line(1,0){32}}
\put(254,50){\line(1,0){32}}
\put(210,54){\line(0,1){32}}
\put(10,50){\circle{8}}
\put(50,50){\circle{8}}
\put(90,50){\circle{8}}
\put(130,50){\circle{8}}
\put(170,50){\circle{8}}
\put(210,50){\circle{8}}
\put(250,50){\circle{8}}
\put(210,90){\circle{8}}
\put(290,50){\circle{8}}
\put(8,30){\shortstack{$\bf 1$}}
\put(48,30){\shortstack{$\bf 2_s$}}
\put(88,30){\shortstack{$\bf 3_1$}}
\put(128,30){\shortstack{$\bf 4_s$}}
\put(168,30){\shortstack{$\bf 5$}}
\put(208,30){\shortstack{$\bf 6_s$}}
\put(248,30){\shortstack{$\bf \bf 4$}}
\put(288,30){\shortstack{$\bf 2_s'$}}
\end{picture}

\noindent Similar considerations can be applied to all finite $SU(2)$
subgroups, all of which are associated with a distinctive extended Dynkin
diagram. For the binary forms of the groups $\m Z_n$, $\m D_n$, $\m T$, $\m O$,
and $\m I$ we have the extended Dynkin diagrams of $A_n$, $D_n$, $E_6$,
$E_7$, and $E_8$, respectively.

\section{$\bs{\mathcal {PSL}_2(7)}$}
Finite groups with complex three-dimensional representations are subgroups of
continuous $SU(3)$. We refer the reader to the previously mentioned
textbook~\cite{oldbook} for the complete list, which contains only one {\em
simple} group, $\mathcal{PSL}_2(7)$.\footnote{Labeled $\Sigma (168)$ in
Refs.~\cite{oldbook,Fairbairn}.} In the remainder of this paper we work out its
representations and discuss some of its properties.   

$\mathcal{PSL}_2(7)$ is the projective special linear group of $(2\times 2)$ matrices over $\mathbb F_7$, the finite Galois field of seven elements.  It contains $168$ elements, and has one complex three-dimensional irrep and its conjugate. It is isomorphic to $\mathcal{GL}_3(2)$, the group of non-singular $(3\times 3)$ matrices with entries in $\mathbb F_2$.  It is therefore an interesting candidate group for explaining the three chiral families. 

As we have seen, it is economical to describe finite groups in terms of their presentation. The
$\mathcal{PSL}_2(7)$ presentation we use is given in terms of two generators $A$ and $B$, as 
\be
<\,A, B\,|\,A^2~=~B^3~=~(AB)^7~=~[A,B]^4~=~1\,>\ ,\ee
where $1$ is the identity element, and $[A,B]=A^{-1}B^{-1}AB$. Written as
$(2\times 2)$ matrices over $\mathbb F_7$, the generators of $\m P \m S \m L_2(7)$ can be taken to be 
\be
A=\pmatrix{0&-1\cr 1&0}\ ,\qquad B~=~\pmatrix{0&-1\cr 1&1}\ ,\ee
so that 
\be
AB=\pmatrix{1&1\cr 0&1}\ ,\qquad [A,B]~=~\pmatrix{1&1\cr 1&2}\ .\ee
The group has six classes

\bea
C^{[1]}_1(1)&& \mbox{ with one element}\nonumber\ ,\\
C^{[2]}_2(A)&& \mbox{  with 21 elements}\nonumber\ ,\\
C^{[3]}_3(B)&& \mbox{  with 56 elements}\nonumber\ ,\\
C^{[4]}_4([A,B])&& \mbox{  with 42 elements}\nonumber\ ,\\
C^{[7]}_5(AB)&& \mbox{  with 24 elements}\nonumber\ ,\\
C^{[7]}_6(AB^2)&& \mbox{  with 24 elements}\nonumber\ ,\eea
where we have indicated  the simplest element of each class in parentheses. It has therefore six irreps, whose properties are summarized in the character table: 

\vskip .3cm
\begin{center}
\begin{tabular}{c|cccccc}
&   \\
{\large$\bs {\m P \m S \m L_2(7)}$}~~& $C^{[1]}_1$& $21C^{[2]}_2(A)$& $56C^{[3]}_3(B)$&$42C^{[4]}_4([A,B])$&$24C^{[7]}_5(AB)$&$24C^{[7]}_6(AB^2)$\\
 &&&&&&   \\
\hline    
&&&&&&   \\
$\chi^{[\bf 1]}_{}$~~& $1$& $1$&$1$&$1$& $1$& $1$\\
&&&&&&   \\
$\chi^{[\bf 3]}_{}$~~& $3$& $-1$& $0$& $1$&$b^{}_7$& $\overline b^{}_7$\\
&&&&&&   \\
$\chi^{[\bf \bar 3]}_{}$~~& $3$& $-1$& $0$& $1$& $\overline b^{}_7$& $b^{}_7$\\
&&&&&&   \\
$\chi^{[\bf 6]}_{}$~~& $6$& $2$& $0$& $0$& $-1$& $-1$\\
&&&&&&   \\
$\chi^{[\bf 7]}_{}$~~& $7$& $-1$& $1$& $-1$& $0$& $0$\\
&&&&&&   \\
$\chi^{[\bf 8]}_{}$~~& $8$& $0$& $-1$& $0$&$1$& $1$\\
\end{tabular}\end{center}

\vskip 1cm
\noindent where we have used the ``Atlas of Finite Groups" \cite{atlas} notation\be
b_7^{}~=~\frac{1}{2}(-1+i\sqrt{7})\ ,\qquad \overline b_7^{}~=~\frac{1}{2}(-1-i\sqrt{7}) \ .\ee

\vskip .3cm
\noindent From the character table, we can directly compute the Kronecker
products of the irreducible representations. The result is:
\vskip .3cm
\begin{center}
{{\begin{tabular}{|c|}
 \hline  \\
~{\bf $\bs {\m P\m S\m L_2(7)}~$  Kronecker Products}\hfill\\
   \\
\hline    
 \\
${\bf 3}~\otimes {\bf 3}~=~{\bf\overline 3}^{}_a~+~{\bf 6}^{}_s\hfill$ \\
${\bf 3}~\otimes {\bf\overline 3}~=~{\bf  1}~+~{\bf 8}\hfill$ \\
\\
$ {\bf 3}~\otimes {\bf 6}~=~{\bf\overline 3}~+~{\bf 7}~+~{\bf 8}\hfill$ \\ 
${\bf \overline 3}~\otimes {\bf 6}~=~{\bf 3}~+~{\bf 7}~+~{\bf 8}\hfill$ \\
${\bf 3}~\otimes {\bf 7}~=~{\bf 6}~+~{\bf 7}~+~{\bf 8}\hfill$ \\
${\bf \overline 3}~\otimes {\bf 7}~=~{\bf 6}~ +~{\bf 7}~ +~{\bf 8}\hfill$ \\
${\bf 3}~\otimes {\bf 8}~=~{\bf 3}~+~{\bf 6}~+~{\bf 7}~+~{\bf 8}\hfill$ \\
${\bf \overline 3}~\otimes {\bf 8}~=~{\bf\overline 3}~ +~{\bf 6}~ +~{\bf 7}~ +~{\bf 8}\hfill$ \\
\\
${\bf 6}~\otimes {\bf 6}~=~({\bf 1}^{}~ +~{\bf 6}^{}~ +~{\bf 6}^{}~ +~{\bf 8}^{})^{}_s~ +~({\bf 7}^{}~ +~{\bf 8}^{})^{}_a\hfill$ \\
${\bf 6}~\otimes {\bf 7}~=~{\bf 3}~+~{\bf\overline 3}~ +~{\bf 6}~ +~{\bf 7}~ +~{\bf 7}~ +~{\bf 8}~ +~{\bf 8}\hfill$ \\
${\bf 6}~\otimes {\bf 8}~=~{\bf 3}~ +~{\bf \overline 3}~ +~{\bf 6}~ +~{\bf 6}~ +~{\bf 7}~ +~{\bf 7}~ +~{\bf 8}~ +~{\bf 8}\hfill$ \\
\\
${\bf 7}~\otimes {\bf 7}~=~({\bf 1}~ +~{\bf 6}~ +~{\bf 6}~ +~{\bf 7}~ +~{\bf 8})^{}_s~ +~({\bf 3}~ +~{\bf \overline 3}~ +~{\bf 7}~ +~{\bf 8})^{}_a\hfill$ \\
${\bf 7}~\otimes {\bf 8}~=~{\bf 3}~ +~{\bf \overline 3}~ +~{\bf 6}~ +~{\bf 6}~ +~{\bf 7}~ +~{\bf 7}~ +~{\bf 8}~ +~{\bf 8}~ +~{\bf 8}\hfill$ \\
\\
${\bf 8}~\otimes {\bf 8}~=~({\bf 1}~ +~{\bf 6}~ +~{\bf 6}~ +~{\bf 7}~ +~{\bf 8}~ +~{\bf 8})^{}_s ~ +~({\bf 3}~ +~{\bf \overline 3}~ +~{\bf 7}~ +~{\bf 7}~+~{\bf 8})^{}_a\hfill$ \\ \\ \hline
\end{tabular}}}\end{center}
\vskip .5cm

\noindent{\bf $\bs{SU(3)}$ Subgroup}
\vskip .3cm

\noindent Since $\mathcal {PSL}_2(7)$ is  a subgroup of $SU(3)$, linear combinations of its six irreducible representations must add up to $SU(3)$ representations. Since the lowest non-trivial representation of both groups is the triplet,  they must be the same. It immediately follows that the antitriplets also correspond to each other (identifying the triplet of one with the antitriplet of the other is a matter of convention).

Consider the product of two $SU(3)$ triplets, which yields ${\bf \bar 3} + {\bf 6}$. Comparison with the Kronecker product of $\m P \m S \m L_2(7)$
shows that the ${\bf 6}$ of $SU(3)$ and the ${\bf 6}$ of $\mathcal{ PSL}_2(7)$ correspond to each other. Similarly, the product of a triplet and an antitriplet shows that the $\bf 8$ of both groups match. This method can be applied to obtain the decompositions of the smallest irreps of $SU(3)$, as shown in the table below:  

\vskip .3cm
\begin{center}
{{\begin{tabular}{|c|}
 \hline  \\
~~$\bs{ SU(3)~ \supset~ \m P \m S \m L_2(7)}$  \hfill\\
   \\
\hline    
 \\
 $(10):~{\bf 3}~=~{\bf 3}\hfill$ \\ 
 $(01):~{\bf \overline 3}~=~{\bf  \overline 3}\hfill$ \\
 $(20):~{\bf 6}~=~{\bf 6}\hfill$ \\
 $(02):~{\bf \overline 6}~=~{\bf 6}\hfill$ \\
$(11):~{\bf 8}~=~{\bf 8}\hfill$ \\ 
$ (30):~{\bf 10}~=~{\bf \overline 3}+{\bf 7}\hfill$ \\
$ (21):~{\bf 15}~=~{\bf 7}+{\bf 8}\hfill$ \\
$ (40):~{\bf 15'}\:\!=~{\bf 1}+{\bf 6}+{\bf 8}\hfill$ \\
$ (05):~{\bf 21}~=~{\bf 3}+{\bf\overline 3}+{\bf 7}+{\bf 8}\hfill$ \\
$ (13):~{\bf 24}~=~{\bf\overline 3}+{\bf 6}+{\bf 7}+{\bf 8}\hfill$ \\
$ (22):~{\bf 27}~=~{\bf 6}+{\bf 6}+{\bf 7}+{\bf 8}\hfill$ \\
\\ \hline
\end{tabular}}}\end{center}
\vskip .5cm

\noindent Although the smallest irreps of the two groups look the same, the differences begin at the level of the sextet. In $SU(3)$ the sextet is a complex representation, while in $\mathcal{PSL}_2(7)$ it is  {\em real} (see later). In  $SU(3)$,  there is a unique invariant made up of three sextets, $\bf 6 \bf 6 \bf 6$; in $\mathcal{PSL}_2(7)$ the Kronecker product shows two separate cubic invariants. A second and more obvious difference is that $SU(3)$ has no seven-dimensional representation.

\section{$\bs{\mathcal{PSL}_2(7)}$ Irreducible Representations}
The purpose of this section is to give explicit matrix realizations for the six irreducible representations of $\mathcal{ PSL}_2(7)$. It is sufficient to derive the matrix expression for its two generators, $A$ and $B$; in the process we will also derive some of the Clebsch-Gordan coefficients.

\subsection{The Triplet Representation}
From the table of Kronecker products, we see that all irreducible representations can be generated from the $\bf 3$; once we compute the $\mathcal{ PSL}_2(7)$ generators in the triplet representation, we can deduce all the others by taking multiple products of the triplet representation. 

We use the presentation and the character table to build the triplet representation. Let $\eta$ be a seventh root of unity, that is
\be
\eta^7_{}~=~1\ .\ee
It is well known (in some circles), that 
\be
b^{}_7~=~\frac{1}{2}(-1+i\sqrt{7})~=~\eta~+~\eta^2~+~\eta^4\ ,\ee
so that, in the triplet representation, the trace of the elements of order seven is a sum of three powers of $\eta$. Folding in the fact that we are interested in an irrep of $(3\times 3)$ unitary matrices with unit determinant, we choose a basis where $AB$ is diagonal, and set 
\be
AB~=~\pmatrix {\eta&0&0\cr 0&\eta^2&0\cr 0&0&\eta^4}\ .\ee
Since $B$ is an element of order $3$, it follows that

$$
A~=~\pmatrix {\eta&0&0\cr 0&\eta^2&0\cr 0&0&\eta^4}B^2\ ,$$
but $B$ is a unitary matrix, that is $B^2=B^\dagger$, so that 

$$
A~=~\pmatrix {\eta&0&0\cr 0&\eta^2&0\cr 0&0&\eta^4}B^\dagger\ .$$
In addition,  $A^2=1$ infers that

$$
B^2\pmatrix {\eta&0&0\cr 0&\eta^2&0\cr 0&0&\eta^4}~=~\pmatrix {\overline\eta&0&0\cr 0&\overline\eta^2&0\cr 0&0&\overline\eta^4}B\ ,$$
which means that the matrix 

$$
\pmatrix {\overline\eta&0&0\cr 0&\overline\eta^2&0\cr 0&0&\overline\eta^4}B\ ,$$
is its own inverse, and therefore hermitian. This restriction allows us to write\be
B~=~\pmatrix{x\eta &a\eta^n&b\eta^m\cr a\eta^{3-n}&y\eta^2&c\eta^p\cr b\eta^{5-m}&c\eta^{6-p}&z\eta^4}\ ,\ee
where $a,b,c,x,y,z$ are real, and $m,n,p$ are integers to be determined. The character table for the triplet yields     

\bean -1&=&\Tr \,A~=~x~+~y~+~z\ ,\\
0&=&\Tr\, B~=~x\eta~+~y\eta^2~+~z\eta^4\ ,\eean
from which  it is straightforward to determine $x,y,z$
\be
x~=~\frac{i}{\sqrt{7}}(\eta^2-\eta^5)\ ,\qquad y~=~\frac{i}{\sqrt{7}}(\eta^4-\eta^3)\ ,\qquad z~=~\frac{i}{\sqrt{7}}(\eta-\eta^6)\ .\ee
It follows that (absorbing $\sqrt{7}$ in $a,b,c$) 
\be
B~=~\frac{i}{\sqrt{7}}\pmatrix{\eta^3-\eta^6 &-ia\eta^n&-ib\eta^m\cr -ia\eta^{3-n}&\eta^6-\eta^5&-ic\eta^p\cr -ib\eta^{5-m}&-ic\eta^{6-p}&\eta^5-\eta^3}\ .\ee
Now

$$
B^\dagger_{}~=~-\frac{i}{\sqrt{7}}\pmatrix{\eta^4-\eta &ia\eta^{n-3}&ib\eta^{m-5}\cr ia\eta^{-n}&\eta-\eta^2&ic\eta^{p-6}\cr ib\eta^{-m}&ic\eta^{-p}&\eta^2-\eta^4}\ .$$
Since $B$ is unitary, $B^\dagger_{}B~=~1\ ,$ we find 

\bean
a(1-\eta+\eta^2-\eta^3)-ibc\eta^{m-p-n}&=&0\ ,\\
b(1+\eta-\eta^3-\eta^4)+iac\eta^{n+p-m}&=&0\ ,\\
c(\eta^2-\eta-\eta^4+\eta^6)-iab\eta^{m-p-n}&=&0\ ,\eean
for the off-diagonal elements, and 

\bean
2-\eta^3-\eta^4+a^2+b^2&=&7\ ,\\
2-\eta-\eta^6+a^2+c^2&=&7\ ,\\
2-\eta^2-\eta^5+b^2+c^2&=&7\ ,\eean
for the diagonal elements. The vanishing of the off-diagonal elements fixes

$$
n-m+p~=~2~\mbox{   mod  }7\ ,$$
and

\bean
bc&=&2a\left(\sin\frac{4\pi}{7}-\sin\frac{6\pi}{7}\right)\ ,\\
ac&=&2b\left(\sin\frac{4\pi}{7}+\sin\frac{2\pi}{7}\right)\ ,\\
ab&=&2c\left(\sin\frac{2\pi}{7}-\sin\frac{6\pi}{7}\right)\ .\eean
These have a sign ambiguity as we can change the sign of any two
variables without affecting the equations. We choose the solutions
\be
a~=~i(\eta-\eta^6)\ ,\qquad b~=~i(\eta^4-\eta^3)\ ,\qquad c~=~i(\eta^2-\eta^5)\ ,\ee
as well as 
\be
n~=~2\ ,\qquad m~=~p~=~4\ .\ee
Assembling these results, we find the explicit form of the generators in the
triplet representation\footnote{The generators
$A(\frac{2\pi}{7},\frac{4\pi}{7})$, $E(0,0)$, $Z(\xi)$ of
Ref.~\cite{Fairbairn} are related to our matrices $A$ and $B$ by
$A(\frac{2\pi}{7},\frac{4\pi}{7})=AB$,~
$E(0,0)=[(AB)^4(AB^2)^2A] \! \cdot \! B \! \cdot \! [(AB)^4(AB^2)^2A]^{-1}$,~
$Z(\xi=e^{\frac{6\pi i}{7}})=A$.}
\bea
B^{[\bf 3]}_{}&=&\frac{i}{\sqrt{7}}\pmatrix{\eta^3-\eta^6 &\eta^3-\eta& \eta-1 \cr \eta^2-1&\eta^6-\eta^5&\eta^6-\eta^2\cr \eta^5-\eta^4&\eta^4-1&\eta^5-\eta^3}\ ,
\\\nonumber &&\\[2mm]
A^{[\bf 3]}_{}&=&\frac{i}{\sqrt{7}}\pmatrix{\eta^2-\eta^5 &\eta-\eta^6& \eta^4-\eta^3 \cr \eta-\eta^6&\eta^4-\eta^3&\eta^2-\eta^5\cr \eta^4-\eta^3&\eta^2-\eta^5&\eta-\eta^6} \ .\eea
The matrix $A$ is manifestly real and symmetric, reproducing Klein's
involution for $A$, which can be found in Klein's original 1878
paper~\cite{klein}. 

The generators in the antitriplet representation, $ A^{[\bf\overline 3]}$ and $ B^{[\bf \overline 3]}$, are simply the complex conjugates of the generators in the triplet. To see this, we note that  $\overline A~=~A^T_{}$ is an involution $ \overline A \overline A~=~A^T_{}A^T_{}~=~(AA)^T_{}~=~1$.
Also, the matrix $\overline B~=~(BB)^T_{}$ obeys
$$
 \overline B~ \overline B~=~B^T_{}\ ,$$
and is of order three as $ \overline B~ \overline B~\overline B~=~B^T_{}\overline B~=~\overline B^\dagger_{}\overline B~=~1$. Hence
\be
A^{[\bf\overline 3]}_{} ~=~\overline A^{[\bf 3]}\ ,\qquad B^{[\bf\overline 3]}_{} ~=~\overline B^{[\bf 3]}_{}\ .\ee

\subsection{The Sextet Representation}
The explicit construction of the sextet from the symmetric product of two triplets allows us to derive at the same time the Clebsch-Gordan coefficients. 

Consider two independent irreps, $\bf 3$ and $\bf 3'$, spanned by $\ket i$ and $ \ket{ i'}$, $i,i'=1,2,3$. We denote the six symmetric states of the sextet $\bf 6$ with a special ket notation as $\vert\,\alpha\,\}$, $\alpha=1,2,...,6$:
\be
 \vert\,\alpha\,\}~=~\sum_{i,j}K^{ij}_{~~\alpha}\ket i\ket {j'}\ ,\ee
where $K^{ij}_{~~\alpha}$ are the Clebsch-Gordan coefficients. Their values
are determined by setting  

$$
\vert\, 1\,\}=\ket  {1} \ket{ 1'}\ ,\quad\vert\, 2\,\}= \ket  {2} \ket{ 2'}\
,\quad\vert\, 3\,\}=\ket  {3} \ket{ 3'}\ , $$
\bea
\vert\, 4\,\}&=&\frac{1}{\sqrt{2}}\left(\ket 1 \ket{ 2'}+~\ket 2 \ket{ 1'}\right)\ ,\cr
\vert \,5\,\}&=&  \frac{1}{\sqrt{2}}\left(\ket 3 \ket{ 1'}+~\ket 1 \ket{ 3'}\right) \ , \cr 
\vert\, 6\,\}&=&   \frac{1}{\sqrt{2}}\left(\ket 2 \ket{ 3'}+~\ket 3 \ket{ 2'}\right)\ \ .\eea
We display the same information in the following  Clebsch-Gordan table 

\vskip .3cm
\begin{center}
$
\begin{array}{c||cccccc}
\multicolumn{1}{c||}{\phantom{\Big|}  {\bf{3}}\otimes{\bf{3}}~} 
& \multicolumn{6}{c}{{\bf{6}} } \\\hline
 \phantom{\Big|}|\: i > \: |\: j' > ~ &  |\: 1 \, \}  &  |\: 2 \, \}  
&  |\: 3 \, \}  &  |\: 4 \, \}  &  |\: 5 \, \}  &  |\: 6 \, \} 
\\\hline\hline &&&&&&\\[-3mm]
\phantom{|}|\: 1 > \: |\: 1' > ~ & 1 & 0 &  0  & 0 & 0 & 0     \\[1mm]  
\phantom{|}|\: 1 > \: |\: 2' > ~ & 0 & 0 &  0  & \frac{1}{\sqrt{2}}  & 0 & 0     \\[1mm]
\phantom{|}|\: 1 > \: |\: 3' > ~ & 0 & 0 &  0  & 0 & \frac{1}{\sqrt{2}}  & 0     \\[1mm] 
\phantom{|}|\: 2 > \: |\: 1' > ~ & 0 & 0 &  0  & \frac{1}{\sqrt{2}}  & 0 & 0     \\[1mm] 
\phantom{|}|\: 2 > \: |\: 2' > ~ & 0 & 1 &  0  & 0 & 0 & 0     \\[1mm] 
\phantom{|}|\: 2 > \: |\: 3' > ~ & 0 & 0 &  0  & 0 & 0 & \frac{1}{\sqrt{2}}     \\[1mm]
\phantom{|}|\: 3 > \: |\: 1' > ~ & 0 & 0 &  0  & 0 & \frac{1}{\sqrt{2}} & 0     \\[1mm] 
\phantom{|}|\: 3 > \: |\: 2' > ~ & 0 & 0 &  0  & 0 & 0 & \frac{1}{\sqrt{2}}     \\[1mm] 
\phantom{|}|\: 3 > \: |\: 3' > ~ & 0 & 0 &  1  & 0 & 0 & 0     
\end{array}
$
\end{center}

\vskip .5cm
In order to get the explicit representations for the generators, we set
\be
A^{[\bf 6]}~=~A^{[\bf 3]}_{}\,A^{[\bf 3]'}_{}\ ,\ee
where $A^{[\bf 3]}_{}$ acts on $\ket i$, and $A^{[\bf 3]'}_{}$ on $\ket {i'}$, that is 
\be
A^{[\bf 6]}\ket  {i} \ket{ j'}~=~A^{[\bf 3]}_{}\ket  {i}A^{[\bf 3]'}_{}\ket{ j'}~=~\sum_{m,n}a^{}_{i\,m}a^{}_{j\,n}\ket  {m} \ket{ n'}\ ,\ee
where $a_{im}$ are the matrix elements of $A$ in the triplet that we have just derived. A tedious but straightforward calculation yields the $(6\times 6)$ matrix 

\vskip .3cm
{\footnotesize
\be 
A_{}^{[\bf 6]}~=~-\frac{2\sqrt{2}}{7}\pmatrix{
\frac{1}{\sqrt{2}}(c^{}_3-1)&\frac{1}{\sqrt{2}}(c^{}_2-1)&\frac{1}{\sqrt{2}}(c^{}_1-1)&c^{}_3-c^{}_1&c^{}_1-c^{}_2&c^{}_2-c^{}_3\cr 
\frac{1}{\sqrt{2}}(c^{}_2-1)&\frac{1}{\sqrt{2}}(c^{}_1-1)&\frac{1}{\sqrt{2}}(c^{}_3-1)&c^{}_2-c^{}_3&c^{}_3-c^{}_1&c^{}_1-c^{}_2\cr 
\frac{1}{\sqrt{2}}(c^{}_1-1)&\frac{1}{\sqrt{2}}(c^{}_3-1)&\frac{1}{\sqrt{2}}(c^{}_2-1)&c^{}_1-c^{}_2&c^{}_2-c^{}_3&c^{}_3-c^{}_1\cr 
c^{}_3-c^{}_1&c^{}_2-c^{}_3&c^{}_1-c^{}_2&\frac{1}{\sqrt{2}}(c^{}_1-1)&\frac{1}{\sqrt{2}}(c^{}_2-1)&\frac{1}{\sqrt{2}}(c^{}_3-1)\cr
c^{}_1-c^{}_2&c^{}_3-c^{}_1&c^{}_2-c^{}_3&\frac{1}{\sqrt{2}}(c^{}_2-1)&\frac{1}{\sqrt{2}}(c^{}_3-1)&\frac{1}{\sqrt{2}}(c^{}_1-1)\cr 
 c^{}_2-c^{}_3&c^{}_1-c^{}_2&c^{}_3-c^{}_1&\frac{1}{\sqrt{2}}(c^{}_3-1)&\frac{1}{\sqrt{2}}(c^{}_1-1)&\frac{1}{\sqrt{2}}(c^{}_2-1)},\ee
}
\vskip .2cm
\noindent  written in terms of 
\be
c^{}_n~=~\cos\left(\frac{2n\pi}{7}\right)\ ,~~n=1,2,3\ , \ee
which satisfy 
\be
c_1+c_2+c_3=-\frac{1}{2}\ .\ee
This enables us to check that the trace

$$
\Tr\, A_{}^{[\bf 6]}~=~-\frac{2}{7}\left(-6+2(c^{}_1+c^{}_2+c^{}_3)\right)~=~2\ ,$$
is consistent with the character table. We note also that 
\be
 s^{}_n~=~\sin\left(\frac{2n\pi}{7}\right)\ ,\qquad s_1+s_2-s_3~=~\frac{\sqrt{7}}{2}\ .\ee

In order to derive the form of $B^{[\bf 6]}$, we first note that the order-seven combination of the generators is the diagonal matrix
\be
A_{}^{[\bf 6]}B_{}^{[\bf 6]}={\rm diag}\,(\eta^2,\,\eta^4,\,\eta,\,\eta^3,\,\eta^5,\,\eta^6)\ .\ee
Since
\be
B_{}^{[\bf 6]}~=~A_{}^{[\bf 6]}\left(A^{[\bf 6]}_{}B_{}^{[\bf 6]}\right)\ ,\ee
we easily find the second generator

\vskip .3cm

$ B^{[\bf 6]}~=~-\frac{2\sqrt{2}}{7}\times$
{\footnotesize
\be\pmatrix{
\frac{\eta^2}{\sqrt{2}}(c^{}_3-1)&\frac{\eta^4}{\sqrt{2}}(c^{}_2-1)&\frac{\eta}{\sqrt{2}}(c^{}_1-1)&\eta^3(c^{}_3-c^{}_1)&\eta^5(c^{}_1-c^{}_2)&\eta^6(c^{}_2-c^{}_3)\cr 
\frac{\eta^2}{\sqrt{2}}(c^{}_2-1)&\frac{\eta^4}{\sqrt{2}}(c^{}_1-1)&\frac{\eta}{\sqrt{2}}(c^{}_3-1)&\eta^3(c^{}_2-c^{}_3)&\eta^5(c^{}_3-c^{}_1)&\eta^6(c^{}_1-c^{}_2)\cr 
\frac{\eta^2}{\sqrt{2}}(c^{}_1-1)&\frac{\eta^4}{\sqrt{2}}(c^{}_3-1)&\frac{\eta}{\sqrt{2}}(c^{}_2-1)&\eta^3(c^{}_1-c^{}_2)&\eta^5(c^{}_2-c^{}_3)&\eta^6(c^{}_3-c^{}_1)\cr 
\eta^2(c^{}_3-c^{}_1)&\eta^4(c^{}_2-c^{}_3)&\eta(c^{}_1-c^{}_2)&\frac{\eta^3}{\sqrt{2}}(c^{}_1-1)&\frac{\eta^5}{\sqrt{2}}(c^{}_2-1)&\frac{\eta^6}{\sqrt{2}}(c^{}_3-1)\cr
\eta^2(c^{}_1-c^{}_2)&\eta^4(c^{}_3-c^{}_1)&\eta(c^{}_2-c^{}_3)&\frac{\eta^3}{\sqrt{2}}(c^{}_2-1)&\frac{\eta^5}{\sqrt{2}}(c^{}_3-1)&\frac{\eta^6}{\sqrt{2}}(c^{}_1-1)\cr 
\eta^2( c^{}_2-c^{}_3)&\eta^4(c^{}_1-c^{}_2)&\eta(c^{}_3-c^{}_1)&\frac{\eta^3}{\sqrt{2}}(c^{}_3-1)&\frac{\eta^5}{\sqrt{2}}(c^{}_1-1)&\frac{\eta^6}{\sqrt{2}}(c^{}_2-1)}.\ee
}
\vskip .2cm
\noindent This matrix is no longer hermitian, but we easily check that it is traceless, as required by the character table.

Although $B^{[\bf 6]}$ and $A^{[\bf 6]}B^{[\bf 6]}$ are complex matrices, their traces are real. It means that $\bf 6$ and $\bf\overline 6$ are equivalent, in the sense that their generators are related by a similarity transformation that is itself a group element (inner automorphism). 

Our basis has the advantage that we can easily read-off the Clebsch-Gordan coefficients, but the quadratic invariant made out of two sextets involves a special matrix. To find it, we start from 
\be
I^{[2]}_{}~=~\sum_{\alpha,\beta}\m C_{}^{\alpha\beta}\vert\,\alpha\,\}\vert\, \beta\,\}\ ,\ee
where $\m C$ is to be determined. Invariance under  transformations  generated by $A$ and $B$ requires, in matrix form, 
\be
\m C~=~(A_{}^{[\bf 6]})^T\m CA_{}^{[\bf 6]}~=~(B_{}^{[\bf 6]})^T\m
CB_{}^{[\bf 6]}\ .\ee \newpage \noindent From  
$$
\m C~=~(A^{[\bf 6]}B^{[\bf 6]})^T\m CA^{[\bf 6]}B^{[\bf 6]}\ ,$$
we deduce that the non-zero elements of the symmetric $\m C$ are $\m C^{15}=\m C^{51}$, $\m C^{24}=\m C^{42}$, and  $\m C^{36}=\m C^{63}$. 
The unknown coefficients are easily determined by using  
$$
A^{[\bf 6]}\m C~=~\m CA^{[\bf 6]}\ ,$$
yielding (up to an overall normalization)
\be
\m C~=~\pmatrix{0&0&0&0&1&0\cr
0&0&0&1&0&0\cr
0&0&0&0&0&1\cr
0&1&0&0&0&0\cr
1&0&0&0&0&0\cr
0&0&1&0&0&0} \ .
\ee
Since $\m C$ is a symmetric real matrix, we can find a basis where the
matrices $A^{[\bf 6]}$ and $B^{[\bf 6]}$ are real. This is achieved by the
similarity transformation generated by the matrix 
\be
\m S~=~\m V \m P\ ,\qquad A_{\rm real}^{[\bf 6]}=\m SA_{}^{[\bf 6]}\m
S^{-1}_{}\ , \qquad B_{\rm real}^{[\bf 6]}=\m SB_{}^{[\bf 6]}\m S^{-1}_{}\ ,\ee
where $\m P$ is a diagonal matrix
\be
\m P~=~{\rm diag}(i\eta^3,\,\eta^6,\,\eta^6,\,-\eta^4,\,i,\,-\eta^4)\ ,\ee
and $\m V$ is a symmetric matrix with equal diagonal elements equal to $1/\sqrt{2}$, and with off-diagonal non-zero matrix elements  
\be
 \m V_{15}=\m V_{24}=\m V_{36}=\frac{i}{\sqrt{2}}\ .\ee
In this basis, $\m C$ becomes proportional to the unit matrix, and the quadratic invariant is trivial. 

\subsection{The Octet Representation}
From the Kronecker product

$$
{\bf 3}\otimes{\bf\overline 3}~=~{\bf 1}+{\bf 8}\ ,$$
it is straightforward to construct the octet representation. The singlet is given by the combination
\be
\frac{1}{\sqrt{3}}\left(\ket  {1} \ket{\bar 1}+~\ket 2 \ket{\bar 2}+~\ket 3 \ket{\bar 3}\right)\ .\ee
The eight states are arranged with the familiar $SU(3)$ labeling,  $\vert\, A\,)$, $A=1,2,...,8$,  with

$$
|\,1\,)~=~\frac{1}{\sqrt{2}}\left(\ket 1\ket{\bar 2}+~\ket 2\ket{\bar 1}\right)\ ,\qquad  |\,2\,)~=~\frac{-i}{\sqrt{2}}\left(\ket 1\ket{\bar 2}-~\ket 2\ket{\bar 1}\right)~,$$

$$
|\,3\,)~=~\frac{1}{\sqrt{2}}\left(\ket 1\ket{\bar 1}-~\ket 2\ket{\bar 2}\right)\ ,$$

$$
|\,4\,)~=~\frac{1}{\sqrt{2}}\left(\ket 1\ket{\bar 3}+~\ket 3\ket{\bar 1}\right)\ ,\qquad  |\,5\,)~=~\frac{-i}{\sqrt{2}}\left(\ket 1\ket{\bar 3}-~\ket 3\ket{\bar 1}\right)~,$$

$$
|\,6\,)~=~\frac{1}{\sqrt{2}}\left(\ket 2\ket{\bar 3}+~\ket 3\ket{\bar 2}\right)\ ,\qquad  |\,7\,)~=~\frac{-i}{\sqrt{2}}\left(\ket 2\ket{\bar 3}-~\ket 3\ket{\bar 2}\right)~,$$
 
\be
|\,8\,)~=~\frac{1}{\sqrt{6}}\left(\ket 1\ket{\bar 1}+~\ket 2\ket{\bar 2}-~2\,\ket 3\ket{\bar 3}\right)~.\ee
This corresponds to expressing the octet states, with  the Gell-Mann matrices acting as Clebsch-Gordan coefficients

\be|\,A\,)~=~\frac{1}{\sqrt{2}} \, \sum_{i,j} \left(\lambda_A\right)_{~\, j}^{\,i}\ket{i}\ket{\bar j}\ .\ee
Using the form of $AB$ in the triplet and antitriplet representations, it is relatively easy to work out the order-seven element, with the result  \\
\vspace{5mm}
\footnotesize
\be
(A^{[{\bf 8}]}_{}B^{[{\bf 8}]}_{}) ~=~
\pmatrix{
 c_1 & s_1 & 0 & 0 & 0 & 0 & 0 & 0\cr
 -s_1 & c_1 & 0 & 0 & 0 & 0 & 0 & 0 \cr
 0 & 0 & 1 & 0 & 0 & 0 & 0 & 0\cr
 0 & 0 & 0 & c_3 & s_3 & 0 & 0 & 0 \cr
 0 & 0 & 0 & -s_3 & c_3 & 0 & 0 & 0\cr
 0 & 0 & 0 & 0 & 0 & c_2 & s_2 & 0\cr
 0 & 0 & 0 & 0 & 0 & -s_2 & c_2 & 0\cr
 0 & 0 & 0 & 0 & 0 & 0 & 0 & 1}\ .
\ee
\vspace{5mm}
\normalsize

\noindent It is straightforward, albeit tedious,  to work out $A$ in the octet.  We find
\footnotesize
\be\hspace{-50mm}
A^{[{\bf 8}]}_{}~=~\frac{1}{7} \, \times \ee
\tiny

$$
\pmatrix{
 2-2 c_1 & 0 & 2 c_1+2 c_2-4 c_3 & 2-2 c_2 & 0 & 2-2 c_3 & 0 & 2 \sqrt{3} c_1-2 \sqrt{3} c_2 \cr
 0 & -2-2 c_1+4 c_2 & 0 & 0 & 2+2 c_2-4 c_3 & 0 & -2+4 c_1-2 c_3 & 0 \cr
 2 c_1+2 c_2-4 c_3 & 0 & -c_1+2 c_2-c_3 & -4 c_1+2 c_2+2 c_3 & 0 & 2 c_1-4 c_2+2 c_3 & 0 & \sqrt{3} c_1-\sqrt{3} c_3 \cr
 2-2 c_2 & 0 & -4 c_1+2 c_2+2 c_3 & 2-2 c_3 & 0 & 2-2 c_1 & 0 & 2 \sqrt{3} c_2-2 \sqrt{3} c_3 \cr
 0 & 2+2 c_2-4 c_3 & 0 & 0 & -2+4 c_1-2 c_3 & 0 & 2+2 c_1-4 c_2 & 0 \cr
 2-2 c_3 & 0 & 2 c_1-4 c_2+2 c_3 & 2-2 c_1 & 0 & 2-2 c_2 & 0 & -2 \sqrt{3} c_1+2 \sqrt{3} c_3 \cr
 0 & -2+4 c_1-2 c_3 & 0 & 0 & 2+2 c_1-4 c_2 & 0 & -2-2 c_2+4 c_3 & 0 \cr
 2 \sqrt{3} c_1-2 \sqrt{3} c_2 & 0 & \sqrt{3} c_1-\sqrt{3} c_3 & 2 \sqrt{3} c_2-2 \sqrt{3} c_3 & 0 & -2 \sqrt{3} c_1+2 \sqrt{3} c_3 & 0 & c_1-2 c_2+c_3}.
$$
\normalsize

\vspace{5mm}
 \noindent This real and symmetric matrix satisfies $(A^{[{\bf 8}]})^2 = 1$. Having determined $A^{[{\bf 8}]}$ as well as $(AB)^{[{\bf 8}]}$, it is easy to obtain the order-three generator 
 $B^{[{\bf 8}]}$, by using 
\be
B^{[{\bf 8}]}~=~A^{[{\bf 8}]}_{}(A^{[{\bf 8}]}_{}B^{[{\bf 8}]}_{})\ .\ee
The quadratic invariant for two octet fields  is given by
\be
I^{[2]} ~=~ \sum_{A,B} ~ \delta^{}_{AB} ~ |A\,)\,|B\,)\ .\ee
Note that we do not distinguish upper and lower indices for this representation since this invariant is diagonal.
\subsection{The Septet Representation}
In order to work out the seven-dimensional representation, we consider the
Kronecker product 

$$
{\bf 3} ~ \otimes ~ {\bf 6}  ~=~ \overline{\bf 3} ~+~ {\bf 7} ~+~ {\bf 8} \ .$$
To avoid confusion, we label the states of the ${\bf 7}$ by $|\,a \succ$, $a=1,2,..., 7$. Since we are in a basis where $AB$ is diagonal in both the $\bf 3$ and the $\bf 6$, these eighteen states 
can be split into linear combinations of eigenstates of the form $|\,i> |\,\alpha ~\}$. For instance those with unit eigenvalue are    
\be
 \mu_1 ~ |\,1>|\,6\,\} ~+~ \mu_2 ~ |\,2>|\,5\,\} ~+~ \mu_3 ~ |\,3>|\,4\,\} 
\ , \ee
with $\mu_i$ to be determined by the orthogonality with
respect to the two states in the octet and the one in the septet. 

This enables us to construct the linear combinations that transform according
to irreps of $\m P \m S \m L_2(7)$, and derive in the process the Clebsch-Gordan coefficients. 
\vskip .2cm
\noindent For the ${\overline{\bf 3}}$, we set
\be
\ket{\bar j}~=~\sum_{i,\alpha} \,  \widetilde K^{ij\,\alpha}_{}\ket i|\,\alpha\}\ ,\ee
and notice that these linear combinations have already been determined, since
\be
 \widetilde K^{ij\,\alpha}_{}~=~ \sum_{\beta}\,K^{ij}_{~~\,\beta} \: \m C_{}^{\beta\alpha}\ .\ee
 
\vskip .2cm
\noindent We express the states of the ${\bf 8}$ as the linear combinations
\be
| {A} \,)~=~\sum_{i,\alpha} \, M^{i\,\alpha}_{~~\,A}\, |i>|\alpha\,\}\ ,\ee
and find the Clebsch-Gordan coefficients. They are to be found in Table A-1.

\vskip .2cm 
\noindent For the ${\bf 7}$, we set
\be
|\,a \succ~=~\sum_{i,\alpha} \, L^{\,i \alpha}_{~~\,a}\, |\,i>|\,\alpha\,\}\ .\ee
These Clebsch-Gordan coefficients can be found in Table A-2.

Although in this basis the phases of the seven states have been chosen as to yield a real ${A}^{[{\bf 7}]}$, $(AB)^{[\bf 7]}$ is not real
\be
(A^{[{\bf 7}] }B^{[{\bf 7}] })= {\rm diag}(1 ,\, \eta,\, \eta^2 ,\,\eta^3,\, \eta^4,\, \eta^5,\,\eta^6) \ ,\ee
corresponding to our ordering of the states. One consequence is that the quadratic invariant is not diagonal (in this basis). Indeed, if we write it as
\be
I^{[2]} ~=~ \sum_{a,b} \, \m D^{ab}_{}\vert a\succ\vert b\succ\ ,\ee
we find the non-zero elements of the symmetric matrix $\m D$ to be
\be
\m D_{}^{11}~=~\m D_{}^{27}~=~\m D_{}^{36}~=~\m D_{}^{45}~=~\m D_{}^{54}~=~\m D_{}^{63}~=~\m D_{}^{72}~=~1\ .\ee
We obtain for the generator  $A^{[{\bf 7}]}$\\

\footnotesize
\be ~\hspace{-30mm}
 A_{}^{[{\bf 7}]} ~=~
\frac{2}{7\sqrt{7}} \, \times 
\ee
\tiny

$$
\pmatrix{
 -\frac{\sqrt{7}}{2} & \sqrt{14} & \sqrt{14} & \sqrt{14} & \sqrt{14} & \sqrt{14} & \sqrt{14}\cr
 \sqrt{14} & s_2+4 s_3 & s_1-4 s_2 & 2 s_1-2 s_2-4 s_3 & -4 s_1-s_3 & 4 s_1+2 s_2+2 s_3 & -2 s_1+4 s_2-2 s_3 \cr
 \sqrt{14} & s_1-4 s_2 & -4 s_1-s_3 & -2 s_1+4 s_2-2 s_3 & s_2+4 s_3 & 2 s_1-2 s_2-4 s_3 & 4 s_1+2 s_2+2 s_3 \cr
 \sqrt{14} & 2 s_1-2 s_2-4 s_3 & -2 s_1+4 s_2-2 s_3 & s_1-4 s_2 & 4 s_1+2 s_2+2 s_3 & s_2+4 s_3 & -4 s_1-s_3 \cr
 \sqrt{14} & -4 s_1-s_3 & s_2+4 s_3 & 4 s_1+2 s_2+2 s_3 & s_1-4 s_2 & -2 s_1+4 s_2-2 s_3 & 2 s_1-2 s_2-4 s_3 \cr
 \sqrt{14} & 4 s_1+2 s_2+2 s_3 & 2 s_1-2 s_2-4 s_3 & s_2+4 s_3 & -2 s_1+4 s_2-2 s_3 & -4 s_1-s_3 & s_1-4 s_2 \cr
 \sqrt{14} & -2 s_1+4 s_2-2 s_3 & 4 s_1+2 s_2+2 s_3 & -4 s_1-s_3 & 2 s_1-2 s_2-4 s_3 & s_1-4 s_2 & s_2+4 s_3}~.
$$
\normalsize
\vskip .2cm 
\noindent Again $B^{[\bf 7]}$ is obtained by multiplying $A^{[\bf 7]}$ with $(AB)^{[\bf 7]}$, since $A^{[\bf 7]}$ is an involution.  

As for the sextet, one can use a similarity transformation \\

\footnotesize
\be\m U ~=~
\frac{1}{\sqrt{2}} 
\pmatrix{
 \sqrt{2} e^{-\frac{i \pi }{4}} & 0 & 0 & 0 & 0 & 0 & 0 \cr
 0 & 1 & 0 & 0 & 0 & 0 & -i \cr
 0 & 0 & 1 & 0 & 0 & -i & 0 \cr
 0 & 0 & 0 & 1 & -i & 0 & 0 \cr
 0 & 0 & 0 & -i & 1 & 0 & 0 \cr
 0 & 0 & -i & 0 & 0 & 1 & 0 \cr
 0 & -i & 0 & 0 & 0 & 0 & 1} \ ,
\ee

\vspace{5mm}
\normalsize

\noindent to make these matrices real:  $A^{[{\bf 7}]}$ does not change,  that is $ A^{[{\bf 7}]}\equiv \m U{A}^{[{\bf 7}]}  \m U^\dagger$, while  $A^{[{\bf 7}]}B^{[{\bf 7}]}$ becomes
\footnotesize

\be \m U {(AB)}^{[{\bf 7}]} \m U^\dagger~=~
\pmatrix{
 1 & 0 & 0 & 0 & 0 & 0 & 0 \cr
 0 & c_1 & 0 & 0 & 0 & 0 & -s_1  \cr
 0 & 0 & c_2 & 0 & 0 & -s_2 & 0  \cr
 0 & 0 & 0 & c_3 & -s_3 & 0 & 0  \cr
 0 & 0 & 0 & s_3 & c_3 & 0 & 0  \cr
 0 & 0 & s_2 & 0 & 0 & c_2 & 0  \cr
 0 & s_1 & 0 & 0 & 0 & 0 & c_1} \ .
\ee
\normalsize

\vskip .3cm
\noindent In this basis, the quadratic invariant will be trivial, since $\m D$ is brought to the unit matrix (up to a normalization factor).
\section{$ \bs{\mathcal{PSL}_2(7)}$ Subgroups}
$\m P \m S \m L_2(7)$ has two maximal subgroups, $\m S_4$, the order $24$ permutation group on four letters, and the order $21$ Frobenius group $\m Z_7\rtimes\m Z_3$, the semi-direct product of $\m Z_7$ and $\m Z_3$.

\vskip .5cm
\noindent \underline{$\bullet~~\bs{ \m P \m S \m L_2(7) \supset \m S_4}$}
\vskip .3cm
\noindent This $24$ element group represents permutations on four objects. Its presentation is  $<\, a\ , b\,|\,  a^4=1\ , b^2=1\ , (ba)^3=1\,>$, and its class structure with representative elements is 
 $\Gamma^{[2]}_2(b)\ ,\Gamma^{[2]}_3(a^2)\ ,\Gamma^{[3]}_4( ba )\ ,\Gamma^{[4]}_5(a)$, with character table 
\begin{center}
\begin{tabular}{c|ccccc}
&   \\
{\large{$\bs {\m S_4}$}}~~&~~$\Gamma^{[1]}_1$&~~~~$6\Gamma^{[2]}_2$&~~~~$3\Gamma^{[2]}_3$&~~~~$8\Gamma^{[3]}_4$&$6\Gamma^{[4]}_5$\\
 & &&&&  \\
\hline    
& &&&&  \\
$\chi^{[\bf 1]}_{}$~~&\hfill $1$&\hfill $1$&\hfill $1$&\hfill $1$&\hfill$1$\\
& &&&&  \\
$\chi^{[\bf 1']}_{}$~~& \hfill$1$& \hfill$-1$&\hfill $1$&\hfill $1$&\hfill$-1$\\
& &&&&  \\
$\chi^{[\bf 2]}_{}$~~&\hfill $2$& \hfill$0$&\hfill $2$& \hfill$-1$&\hfill$0$\\
& &&&&  \\
$\chi^{[\bf 3_1]}_{}$~~&\hfill $3$&\hfill $1$&\hfill $-1$&\hfill $0$&\hfill$-1$\\
& &&&&  \\
$\chi^{[\bf 3_2]}_{}$~~&\hfill $3$&\hfill $-1$&\hfill $-1$&\hfill $0$&\hfill$1$\\
\end{tabular}\end{center}
\vskip .2cm
Its  generators can  be expressed as $(2\times 2)$ matrices over $\mathbb F_7$; for instance
\be
a=\pmatrix{1&1\cr 1&2}\ ,\qquad b~=~\pmatrix{4&4\cr 1&3}\ ,\ee
which can be expressed in terms of the $\m P \m S \m L_2(7)$ generators as, 
 \be
a ~=~ [\,A,B\,]\ ,\qquad b~= ~A (AB_{}^2)^3 (AB)_{}^3\ .\ee
Its irreducible representations obey the following Kronecker products, see
\cite{S4} for details and the explicit Clebsch-Gordan
coefficients.

\vskip .5cm
\begin{center}
\begin{tabular}{|c|}
 \hline  \\
~{\bf$\bs{\m S_4}$  Kronecker Products}\hfill\\
   \\
\hline    
 \\
 ${\bf 1'}\otimes {\bf 1'}~=~{\bf 1}\hfill$ \\[1mm]

${\bf 2}\;\otimes{\bf 1'}~=~{\bf 2}\hfill$ \\[1mm] 
$ {\bf 3_1}\otimes{\bf 1'}\!~=~{\bf 3_2}\hfill$ \\[1mm]
$  {\bf 3_2}\otimes{\bf 1'}\!~=~{\bf 3_1}\hfill$ \\[1mm]

$  {\bf 2}\:\otimes\,{\bf 2}\:~=~({\bf 1} + {\bf 2})^{}_s + ({\bf 1'})^{}_a\hfill$ \\[1mm]

 ${\bf 2}\,\otimes{\bf 3_1}~=~{\bf 2}\;\otimes\:{\bf 3_2}~=~{\bf 3_1}+{\bf 3_2}\hfill$ \\[1mm]

 ${\bf 3_1}\!\otimes{\bf 3_1}\!~=~{\bf 3_2}\otimes{\bf 3_2}~=~({\bf 1}+{\bf 2}+{\bf 3_1})^{}_s+({\bf 3_2})_a^{}\hfill$ \\[1mm]

 ${\bf 3_1}\!\otimes{\bf 3_2}\!~=~{\bf 1'}+{\bf 2}+{\bf 3_1}+{\bf 3_2}\hfill$ \\
\\ \hline
\end{tabular}
\end{center}
\vskip .5cm

\vskip .5cm
\noindent  $\underline{\bullet~~\bs{\m P \m S \m L_2(7)\supset \m Z_7\rtimes\m Z_3}}$
\vskip .3cm
\noindent This is the Frobenius group of order $21$ with elements of order three and seven.  It has the presentation  $< c\, , d\,|\,  c^7=d^3=1\, , \,d^{-1}cd=c^4\,>$, with classes  
 $\Theta^{[3]}_2(d )\ ,\Theta^{[3]}_3(d^2)\ , \Theta^{[7]}_4(c)\ ,\Theta^{[7]}_5(c^3)$. Its character table reads
\vskip .3cm
\begin{center}
{{\begin{tabular}{c|ccccc}
&   \\
{\large{$\bs{\m Z_7\rtimes \m Z_3}$}}~~&~~$\Theta^{[1]}_1$& $7\Theta^{[3]}_2$& $7\Theta^{[3]}_3$& $3\Theta^{[7]}_4$& $3\Theta^{[7]}_5$\\
 & &&&&  \\
\hline    
& &&&&  \\
$\chi^{[\bf 1]}_{}$~~& $1$& $1$& $1$& $1$& $1$\\
& &&&&  \\
$\chi^{[\bf 1']}_{}$~~& $1$& $e^{2\pi i/3}$& $e^{4\pi i/3}$& $1$& $1$\\
& &&&&  \\
$\chi^{[\bf \overline 1']}_{}$~~& $1$& $e^{4\pi i/3}$& $e^{2\pi i/3}$& $1$& $1$\\
& &&&&  \\
$\chi^{[\bf 3]}_{}$~~& $3$& $0$& $0$& $b_7^{}$& $\overline b_7^{}$\\
& &&&&  \\
$\chi^{[\bf \overline 3]}_{}$~~& $3$& $0$& $0$& $\overline b_7^{}$&$b_7^{}$\\
\end{tabular}}}\end{center}
\vskip .5cm
 Its generators can be  expressed as $(2\times 2)$ matrices over $\mathbb F_7$,
\be
c=\pmatrix{1&1\cr 0&1}\ ,\qquad d~=~\pmatrix{4&0\cr 0&2}\ ,\ee
projectively defined. In terms of the $\mathcal{ PSL}_2(7)$ generators, they are\be
c=AB\ ,\qquad d~=~AB(AB^2)^2(AB)^2AB^2\ .\ee
We note here for completeness the Kronecker products of its irreducible representations. First 

$$
{\bf 3}\otimes {\bf 3}={\bf 3}_s+{\bf\overline 3}_s+{\bf\overline 3}_a\ ,$$
with the Clebsch-Gordan decompositions

$$({\bf 3}\otimes {\bf 3})_s^{}~~\longrightarrow~~{\bf 3}:~~\cases{\ket 3\ket{3'}\cr \ket 1\ket{1'}\cr\ket 2\ket{2'}}\ ;\qquad {\bf\overline 3}:~~\cases{\frac{1}{\sqrt{2}}\left(\ket 3\ket{2'}+~\ket2\ket{3'}\right)\cr\frac{1}{\sqrt{2}}\left(\ket 1\ket{3'}+~\ket 3\ket{1'}\right)\cr\frac{1}{\sqrt{2}}\left(\ket 2\ket{1'}+~\ket1\ket{2'}\right)}\ ;$$
\be({\bf 3}\otimes {\bf 3})_a^{}~~\longrightarrow~~ {\bf\overline 3}:~~\cases{\frac{1}{\sqrt{2}}\left(\ket 3\ket{2'}-~\ket2\ket{3'}\right)\cr\frac{1}{\sqrt{2}}\left(\ket 1\ket{3'}-~\ket 3\ket{1'}\right)\cr\frac{1}{\sqrt{2}}\left(\ket 2\ket{1'}-~\ket1\ket{2'}\right)}\ .\ee
The further product 

$$
{\bf 3}\otimes {\bf\overline 3}={\bf 1}+{\bf 1'}+{\bf\overline 1'}+{\bf 3}+{\bf\overline 3}\ ,$$
yields
\be
{\bf 3}\otimes {\bf\overline 3}~~\longrightarrow~~{\bf 3}:~~\cases{\ket 2\ket{\bar 1'}\cr \ket 3\ket{\bar 2'}\cr\ket 1\ket{\bar 3'}}\ ;\qquad {\bf\overline 3}:~~\cases{\ket 1\ket{\bar 2'}\cr \ket 2\ket{\bar 3'}\cr\ket 3\ket{\bar 1'}}\ ,\ee
as well as the one-dimensional representations

\bea
&&{\bf 1{\phantom{'}}}:~~\frac{1}{\sqrt{3}}\,\left( \ket 1\ket{\bar 1'}+ ~\ket
  2\ket{\bar 2'}+ ~ \ket 3\ket{\bar 3'}\right)\ , \nonumber \\ 
&&{\bf 1'}:~~\frac{1}{\sqrt{3}}\,\left( \ket 1\ket{\bar 1'}+ ~\omega^2_{}\ket
 2\ket{\bar 2'}+ ~\omega\:\ket 3\ket{\bar 3'}\right)\ ,~~~~~~~~~~~~~\\
&&{\bf \overline 1'}:~~\frac{1}{\sqrt{3}}\,\left( \ket 1\ket{\bar 1'}+ ~\omega\:\ket  2\ket{\bar 2'}+~ \omega^2_{}\ket 3\ket{\bar 3'}\right)\ ,\nonumber
\eea
where $\omega=\exp(2i\pi/3)$. Finally, we note that 

$$
{\bf 3}\otimes {\bf 1'}={\bf 3}\ ,\qquad 
{\bf 3}\otimes {\bf\overline 1'}~=~{\bf 3}\ , \qquad 
{\bf 1'}\otimes {\bf 1'}={\bf \overline 1'}\ ,\qquad 
{\bf 1'}\otimes {\bf \overline 1'}={\bf 1}\ .
$$
The relevant information is summarized in the following table.

\vskip .3cm
\begin{center}
{{\begin{tabular}{|c|}
 \hline  \\
~{\bf$\bs{\m Z_7\rtimes \m Z_3}$  Kronecker Products}\hfill\\
   \\
\hline    
 \\
 ${\bf 1'}\otimes {\bf 1'}~=~{\bf \overline 1'}\hfill$ \\ ${\bf 1'}\otimes {\bf \overline1'}~=~{\bf 1}\hfill$ \\
${\bf 3}\:\otimes\!\:{\bf 1'}~=~{\bf 3}\hfill$ \\ 
$ {\bf 3}\:\otimes\!\:{\bf \overline 1'}~=~{\bf 3}\hfill$ \\
 ${\bf 3}\,\otimes\,{\bf 3}\:~=~({\bf 3}+{\bf\overline 3})^{}_s+{\bf \overline 3}_a^{}\hfill$ \\
 ${\bf 3}\,\otimes\,{\bf \overline 3}~\:=~{\bf 1}+{\bf 1'}+{\bf \overline 1'}+{\bf 3}+{\bf \overline3}\hfill$ \\\
\\ \hline
\end{tabular}}}\end{center}
\vskip .4cm

\noindent{\bf Embeddings}
\vskip .3cm

\noindent Invariance under group operations  can be ``spontaneously" broken down to
subinvariances by assigning "vacuum values" to components of fields belonging
to irreps of $\m P\m S\m L_2(7)$ which contain the singlet representation of
the unbroken subgroup.  This requires the decompositions of the irreducible
representations of $\m P\m S\m L_2(7)$ in terms of those of its subgroups, notably  $\m Z_7\rtimes \m Z_3$ and $\m S_4$. We first discuss a general technique which relies on the way its classes contain elements of the subgroup. 

Let $\m G$ be a group of order $n$, with irreps $\mathfrak R^{[\alpha]}$, of dimension $D_\alpha$ and characters $\Xi^{[\alpha]}$. Let $\m K$ be one of its  subgroups,  of order $k$, with irreps  $\mathfrak T^{[a]}$, of dimensions $d_a$ and characters $\chi^{[a]}$. We set 
\be
\mathfrak R^{[\alpha]}_{}~=~\sum_a f^{[\alpha]}_{~~a}\,\mathfrak T^{[a]}_{}\ .\ee
The embedding coefficients $f^{\,[\alpha]}_{~~a}$ are  positive integers, subject to the dimension constraints, 
\be
D^{}_\alpha~=~\sum_a f^{[\alpha]}_{~~a}\,d_{a}^{}\ .\ee
The embedding coefficients  are simply determined from the way which $\m K$ classes fit in the $\m G$ classes:
\begin{itemize}  

\item If the class  $C_i$ of $\m G$ contains elements of $\m K$ which live in its {\em different} classes $\Gamma_{i_m}$, then   
\be
\sum _\alpha f^{[\alpha]}_{~~b}\,\Xi^{[\alpha]}_i~=~\left(\frac{n}{k}\right)\sum_{i_m}\frac{k^{}_{i_m}}{n_i}\chi^{[b]}_{i_m}\ ,\ee
where the sum is over the classes of $\m K$ which are contained in the $C_i$ class of $\m G$. 

\item If the class $C_\perp$ with character $\Xi^{[\alpha]}_\perp$ has no element in $\m K$, a similar reasoning leads to 
\be
\sum _\alpha f^{[\alpha]}_{~~b}\,\Xi^{[\alpha]}_\perp~=~0\ .\ee
\end{itemize}
We now apply these formul{\ae} to the two maximal subgroups.
\vskip .5cm
\noindent \underline{$\bullet~~\bs{\m P\m S\m L_2(7)\supset \m S_4}$}
\vskip .3cm 
\noindent The class structure is 
\be
C^{[2]}_2\supset \Gamma^{[2]}_2+\Gamma^{[2]}_3\ ,\qquad C^{[3]}_3\supset \Gamma^{[3]}_4\ ,\qquad C^{[4]}_4\supset \Gamma^{[4]}_5\ ,\ee
while $C^{[7]}_5$ and $C^{[7]}_6$ contain no $\m S_4$ classes. Hence we find 

{
\be\pmatrix{7\chi^{[a]}_1\cr 2\chi^{[a]}_2+\chi^{[a]}_3\cr \chi^{[a]}_4\cr \chi^{[a]}_5\cr 0\cr 0}=\pmatrix{1&3&3&\hfill 6&\hfill 7&\hfill 8\cr 
1&-1&-1&\hfill 2&\hfill -1&\hfill 0\cr 1&0&0&\hfill 0&\hfill 1&\hfill -1\cr1&1&1&\hfill 0&\hfill -1&\hfill 0\cr 1&b_7^{}&\overline b_7^{}&\hfill -1&\hfill 0&\hfill 1\cr 1&\overline b_7^{}&b_7^{}&\hfill -1&\hfill 0&\hfill 1}
\pmatrix{f^{[\bf 1]}_a\cr 
f^{[\bf 3]}_a\cr f^{[\bf \overline 3]}_a\cr f^{[\bf 6]}_a\cr f^{[\bf 7]}_a\cr f^{[\bf 8]}_a} \ ,
\ee}

\noindent valid for each $[a]$ irrep of $\m S_4$. The embedding coefficients
are obtained by inverting this matrix and inserting  the characters of the
subgroup. We find 

\vskip .3cm
\begin{center}
{{\begin{tabular}{|c|}
 \hline  \\
~$\bs{\m P\m S\m L_2(7)\supset \m S_4}$  \hfill\\
   \\
\hline    
 \\
 ${\bf 3}~=~{\bf 3_2}\hfill$ \\ 
 ${\bf \overline 3}~=~{\bf  3_2}\hfill$ \\
 ${\bf 6}~=~{\bf 1}+{\bf 2}+{\bf  3_1}\hfill$ \\
${\bf 7}~=~{\bf 1'}+{\bf 3_1}+{\bf  3_2}\hfill$ \\ 
$ {\bf 8}~=~{\bf 2}+{\bf 3_1}+{\bf 3_2}\hfill$ \\
\\ \hline
\end{tabular}}}\end{center}
\vskip .5cm

\noindent We see that spontaneous breakdown to $\m S_4$ can be achieved with
the sextet representation. Adopting our basis for the sextet representation, the singlet vev is $(\sqrt{2},\sqrt{2}\eta,\sqrt{2}\eta^3,b^{}_7\eta^4,b^{}_7\eta^5,b^{}_7\eta^2)$. 

Once broken, $\m S_4$ can break to $\m A_4$, its maximal subgroup. We note here for convenience its character table

\vskip .3cm
\begin{center}
\begin{tabular}{c|cccc}
&   \\
{\large$ \bf {\m A_4}$}~~~~&~~$C^{[1]}_1$&~~~~$4C^{[3]}_2$&~~~~$4C^{[3]}_3$&~~~~$3C^{[2]}_4$\\
 & &&&  \\
\hline    
& &&&  \\
$\chi^{[\bf 1]}_{}$~~& $1$& $1$& $1$& $1$\\
& &&&  \\
$\chi^{[\bf 1']}_{}$~~& $1$& $e^{2\pi i/3}_{}$& $e^{4\pi i/3}_{}$& $1$\\
& &&&  \\
$\chi^{[\bf \overline 1']}_{}$~~& $1$& $e^{4\pi i/3}_{}$& $e^{2\pi i/3}_{}$& $1$\\
& &&&  \\
$\chi^{[\bf 3]}_{}$~~& $3$& $0$& $0$& $-1$\\
\end{tabular}\end{center}
\vskip .2cm

\noindent The Kronecker products and the decompositions of the $\m S_4$
representations into those of $\m A_4$ are given as follows. Note that a
vev for the ${\bf 1'}$ of  $\m S_4$ breaks this group down to $\m A_4$.

\vskip .5cm
\begin{center}
\begin{tabular}{ccc}
\begin{tabular}{|c|}
 \hline  \\
~{\bf$\bs{\m A_4}$  Kronecker Products}\hfill\\
   \\
\hline    
 \\
 ${\bf 1'}\otimes {\bf 1'}~=~{\bf \overline 1'}\hfill$ \\
 ${\bf 1'}\otimes {\bf \overline 1'}~=~{\bf 1}\hfill$ \\
$ {\bf 3}\;\otimes{\bf 1'}~=~{\bf 3}\hfill$ \\
$  {\bf 3}\;\otimes{\bf 3}\;~=~{\bf 1} +{\bf 1'} +{\bf \overline 1'} +{\bf 3} + {\bf 3} \hfill$ \\
\\ \hline
\end{tabular}
&~~~~~~&
\begin{tabular}{|c|}
 \hline  \\
~$\bs{\m S_4 \supset \m A_4}$  \hfill\\
   \\
\hline    
 \\
 ${\bf 1'}~=~{\bf 1}\hfill$ \\ 
 ${\bf 2}\;~=~{\bf  1'} + {\bf \overline 1'} \hfill$ \\
 ${\bf 3_1}\;\!=~{\bf 3} \hfill$ \\
${\bf 3_2}\;\!=~{\bf 3} \hfill$ \\ 
\\ \hline
\end{tabular} 
\end{tabular}
\end{center}
\vskip .5cm

\vskip .5cm
\noindent  \underline{$\bullet~~\bs{\m P\m S\m L_2(7)\supset \m Z_7\rtimes\m Z_3}$}
\vskip .3cm
\noindent The Frobenius group has only elements of order three and seven. Hence $C^{[2]}_2$ and $C^{[4]}_4$, with elements of order two and four respectively,  contain no elements of $\m Z_7\rtimes\m Z_3$. On the other hand, we find 
\be
C^{[3]}_3\supset \Theta^{[3]}_2+\Theta^{[3]}_3\ ,\qquad C^{[7]}_5\supset \Theta^{[7]}_4\ ,\qquad C^{[7]}_6\supset \Theta^{[7]}_5\ .\ee
Applying our formul{\ae} to these considerations yields the matrix equation
{
\be\pmatrix{8\chi^{[a]}_1\cr 0\cr \chi^{[a]}_2+\chi^{[a]}_3\cr 0\cr \chi^{[a]}_4\cr \chi^{[a]}_5}=\pmatrix{\hfill 1&\hfill 3&\hfill 3&\hfill 6&\hfill 7&\hfill 8\cr 
1&-1&-1&\hfill 2&\hfill -1&\hfill 0\cr \hfill 1&\hfill 0&\hfill 0&\hfill 0&\hfill 1&\hfill -1\cr\hfill 1&\hfill 1&\hfill 1&\hfill 0&\hfill -1&\hfill 0\cr 1&\hfill b_7^{}&\hfill \overline b_7^{}&\hfill -1&\hfill 0&\hfill 1\cr \hfill 1&\hfill \overline b_7^{}&\hfill b_7^{}&\hfill -1&\hfill 0&\hfill 1}
\pmatrix{f^{[\bf 1]}_a\cr 
f^{[\bf 3]}_a\cr f^{[\bf \overline 3]}_a\cr f^{[\bf 6]}_a\cr f^{[\bf 7]}_a\cr f^{[\bf 8]}_a} \ ,
\ee}
 
 \noindent valid for each irrep of $\m Z_7\rtimes \m Z_3$. This determines the
embedding coefficients, with the following result. 

\vskip .3cm
\begin{center}
{{\begin{tabular}{|c|}
 \hline  \\
~$\bs {\m P\m S\m L_2(7)\supset \m Z_7\rtimes \m Z_3}$ \hfill\\
   \\
\hline    
 \\
 ${\bf 3}~=~{\bf 3}\hfill$ \\ 
 ${\bf \overline 3}~=~{\bf \overline 3}\hfill$ \\
 ${\bf 6}~=~{\bf 3}+{\bf \overline 3}\hfill$ \\
${\bf 7}~=~{\bf 1}+{\bf 3}+{\bf \overline 3}\hfill$ \\ 
$ {\bf 8}~=~{\bf 1'}+{\bf \overline 1'}+{\bf 3}+{\bf\overline 3}\hfill$ \\
\\ \hline
\end{tabular}}}\end{center}
\vskip .5cm
The singlet of the subgroup appears only in the septet, so that only a septet field can spontaneously break to this subgroup. In our representation, the septet vev is simply $(1,0,0,0,0,0,0)$. The only subgroups of the Frobenius group are $\m Z_7$ and $\m Z_3$, so that this breaking chain is rather simple. 


\section{Invariants}
Since we have in mind applications to field theory, we represent the different irreducible representations in terms of fields: 

\begin{itemize}

\item Triplet fields:    $\varphi^{}_i ,\varphi_j'$, \dots ; antitriplets fields:  $\overline\varphi^{i}_{}\ ,{\overline\varphi'}^{j}_{}$, \dots , $i,j=1,2,3$;

\item Sextet fields:  $\chi^{}_\alpha ,\chi'_\beta$, \dots,  with $\alpha,\beta=1,2,\dots, 6$; 

\item Septet fields:  $\psi_a^{} , \psi'_b$, \dots,  with $a,b=1,2,\dots, 7$; 

\item Octet fields: $\Sigma_A ,\Sigma'_B$ , \dots, $A,B=1,2\dots,8$. 

\end{itemize}
\vskip .2cm
\noindent{\bf Quadratic Invariants }
\vskip .2cm
\noindent We have already built all possible quadratic invariants

$$
\varphi_i^{}\,{\overline\varphi'}^{i}_{}\ ; \quad \chi^{}_\alpha\,\chi'_\beta\m C_{}^{\alpha\beta}\ ;\quad \psi_a^{}\, \psi'_b\m D^{ab}_{}\ ;\quad\Sigma_A\,\Sigma'_B\delta^{}_{AB}\ .$$
Note that in our bases, the $\m C$ and $\m D$ matrices are used to raise and lower indices in the $\bf 6$ and $\bf 7$, respectively.\footnote{We could have chosen bases where $\m C$ and $\m D$ are the unit matrices, but  the Clebsch-Gordan coefficients would have been more complicated.}
Note that from now on, we always sum over repeated indices.

\newpage

\vskip .2cm 
\noindent{\bf Cubic Invariants }
\vskip .2cm
\noindent Cubic invariants which contain triplets and antitriplets are given by
\bea
({\bf 3}\otimes {\bf 3})_a&=&{\bf\overline 3}~~~~\longrightarrow~~ \epsilon_{}^{ijk}\,\,\varphi^{}_i\,\varphi_j'\,\varphi_k''\ ,\cr
({\bf 3}\otimes {\bf 3})_s&=&{\bf 6}~~~~\longrightarrow~~ K^{ij}_{~~\alpha}\,\m C^{\alpha\beta}_{}\,\varphi^{}_i\,\varphi_j'\,\chi^{}_\beta\ ,\cr
{\bf 3}\otimes {\bf 6}&=&{\bf 7}~~~~\longrightarrow~~ L^{\,i \alpha}_{~~b}\,\m D^{ab}_{}\,   \varphi^{}_i\, \chi^{}_\alpha\, \psi^{}_a\ ,\cr
{\bf 3}\otimes {\bf 6}&=&{\bf 8}~~~~\longrightarrow~~ M^{i\,\alpha}_{~~\,A} \,\,  \varphi^{}_i \,\chi^{}_\alpha\, \Sigma^{}_A\ ,\cr
{\bf 3}\otimes {\bf\overline 3}&=&{\bf 8}~~~~\longrightarrow~~ \left(\lambda^{}_{A}\right)^{\,i}_{~  j} \varphi^{}_i \,\overline\varphi^{j}_{}\, \Sigma^{}_A\ ,\cr
{\bf 3}\otimes {\bf 7}&=&{\bf 7}~~~~\longrightarrow~~ N^{\,i a}_{~~\,c}\,\m D^{cb}_{}\,   \varphi^{}_i\, \psi^{}_a\, \psi'_b\ ,\cr
{\bf 3}\otimes {\bf 7}&=&{\bf 8}~~~~\longrightarrow~~ P^{\,i a}_{~~\,A}\,   \varphi^{}_i\, \psi^{}_a\, \Sigma^{}_A\ ,\cr
{\bf 3}\otimes {\bf 8}&=&{\bf 8}~~~~\longrightarrow~~ Q^{\,i }_{~AB}\,   \varphi^{}_i\,  \Sigma^{}_A\Sigma'_B\ .\eea

\noindent $N^{\,i a}_{\,~~c}$, $P^{\,i a}_{\,~~A}$, and $Q^{\,i}_{\,~AB}$ are
the Clebsch-Gordan coefficients for the products ${\bf 3} \otimes {\bf 7} =
{\bf 7}$,  ${\bf 3} \otimes  {\bf 7} = {\bf 8}$, and ${\bf 3} \otimes 
{\bf 8} = {\bf 8}$, respectively. They are listed in the tables of Appendix
A. Applying our rules for raising and lowering indices, one can easily derive
other Clebsch-Gordan coefficients, e.g. 

\be
K^{ij}_{~~\alpha}\,\m C^{\alpha\beta}\, \varphi^{}_i \,\chi_\beta^{}\ ,\ee

\noindent has a free upper index $j$, and thus transforms as an antitriplet ${\bf \overline 3}$. 

If we include the other representations, we can build many more cubic invariants, as we can infer from the table of Kronecker products.  As an example, we see that there are two cubic invariants made out of one octet, and three if we consider several octets; as we mentioned earlier, there are {\em two} cubic invariants made out of {\em one} sextet, etc... . 

The list of possible invariants one can construct out of several irreps grows
very quickly. In the following we restrict ourselves to discuss the
invariants which can be constructed out of triplets and antitriplets only.

\vskip .4cm
\noindent{\bf Klein's Quartic Invariant }
\vskip .2cm
 
\noindent With only one triplet, there is a unique quartic invariant

\be
I^{[4]}_{}~\equiv~\frac{1}{2\sqrt{2}}\left(\varphi^{}_i\, \varphi_j^{}K^{ij}_{~~\alpha}\right)\m C^{\alpha\beta}_{}\left(\varphi^{}_k \varphi_l^{}K^{kl}_{~~\beta}\right)\ .\ee

\noindent In terms of components,

\be
I^{[4]}_{}~=~(\varphi^{}_1)^3\varphi_3^{}+(\varphi^{}_3)^3\varphi_2^{}+(\varphi^{}_2)^3\varphi_1^{}\
.\ee

\noindent This quartic invariant was found by Felix Klein, and the equation

\be
I^{[4]}_{}~=~0\ ,\ee

\noindent defines {\em Klein's Quartic Curve}.  Obviously,
$\mathcal{PSL}_2(7)$ acts as the group of conformal transformations of the
quartic into itself.  Klein's quartic can be parametrized in terms of one
complex variable $w$,  with
 
\be
\varphi^{}_1~=~A^3_{}B\ ,\quad \varphi^{}_2~=-2^{1/7}wB^3_{}\ ,\quad
\varphi^{}_3~=~2^{3/7}w^3_{}A\ ,\ee
\vskip -0.1cm
\be A(w)~=~(1+w^7)^{1/7}_{}\ ;\qquad B(w)~=~(1-w^7)^{1/7}_{}\ .\ee

\vskip .2cm
\noindent It can be shown to be a Riemann surface of genus $3$ of constant negative curvature. It is mathematically special, since it has the maximum number of symmetries allowed by its genus. 

\vskip .4cm
\noindent{\bf Higher Order Invariants }
\vskip .2cm
\noindent By taking the determinant of the $(3\times 3)$ matrix
\be
M^{[3]}_{ij}~=~ \frac{\partial^2 I^{[4]}_{}}{\partial \varphi_i\partial\varphi_j}, \ee
Klein found the sixth-order invariant (the Hessian)
\be
I^{[6]}_{}~=~-\frac{1}{54}\det M^{[3]}_{}\ .\ee

\vskip .2cm
\noindent He then formed the $(4\times 4)$ matrix 

\vskip .2cm
\be M^{[4]}_{}=\pmatrix{
\frac{\partial^2 I^{[4]}_{}}{\partial \varphi^2_1}&\frac{\partial^2 I^{[4]}_{}}{\partial \varphi_1\partial\varphi_2}&\frac{\partial^2 I^{[4]}_{}}{\partial \varphi_1\partial\varphi_3}&\frac{\partial I^{[6]}_{}}{\partial \varphi_1}\cr
\frac{\partial^2 I^{[4]}_{}}{\partial \varphi_2\partial\varphi_1}&\frac{\partial^2 I^{[4]}_{}}{\partial \varphi^2_2}&\frac{\partial^2 I^{[4]}_{}}{\partial \varphi_2\partial\varphi_3}&\frac{\partial I^{[6]}_{}}{\partial \varphi_2}\cr
\frac{\partial^2 I^{[4]}_{}}{\partial \varphi_3\partial\varphi_1}&\frac{\partial^2 I^{[4]}_{}}{\partial \varphi_3\partial\varphi_2}&\frac{\partial^2 I^{[4]}_{}}{\partial \varphi^{2}_3}&\frac{\partial I^{[6]}_{}}{\partial \varphi_3}\cr
\frac{\partial I^{[6]}_{}}{\partial \varphi_1}&\frac{\partial I^{[6]}_{}}{\partial \varphi_2}&\frac{\partial I^{[6]}_{}}{\partial \varphi_3}&0
} ,\ee

\vskip .2cm
\noindent to find the invariant of order $14$, 
\be
I^{[14]}_{}~=~\frac{1}{9}\det M^{[4]}_{}\ .\ee

\vskip .2cm
\noindent Finally, there is the invariant of order $21$
\be
I^{[21]}_{}~=~-\frac{1}{14}\det K\ ,\ee
where
\be
K~=~\pmatrix{\frac{\partial I^{[4]}}{\partial\varphi_1}&\frac{\partial
    I^{[4]}}{\partial\varphi_2}&\frac{\partial I^{[4]}}{\partial\varphi_3}\cr 
\frac{\partial I^{[6]}}{\partial\varphi_1}&\frac{\partial
    I^{[6]}}{\partial\varphi_2}&\frac{\partial I^{[6]}}{\partial\varphi_3}\cr 
\frac{\partial I^{[14]}}{\partial\varphi_1}&\frac{\partial
  I^{[14]}}{\partial\varphi_2}&\frac{\partial I^{[14]}}{\partial\varphi_3}}\
.\ee

\newpage
\vskip .2cm 
\noindent{\bf Extreme Values of Klein's Invariant}
\vskip .2cm

\noindent In order to find the configurations which extremize Klein's
invariant, consider the following $\m P\m S\m L_2(7)$-invariant potential
($m^2$ and $\lambda$ are positive)
\be
V ~=~ -m^2 \cdot \sum_i \varphi_i^\dagger\varphi_i^{} 
~+~ \lambda \cdot \left(\sum_i \varphi_i^\dagger\varphi_i^{}\right)^{\!\!2}  
+~ \kappa \cdot (\varphi_1^3 \varphi_3^{} + \varphi_3^3 \varphi_2^{} +
\varphi_2^3 \varphi_1^{} ~+~ \mathrm{c.c.}) \ . 
 \ee

\vspace{2mm}

\noindent Setting 
\be \pmatrix{\varphi_1 \cr \varphi_2 \cr \varphi_3} 
~=~ r \cdot \pmatrix{\phi_1 \cr \phi_2 \cr \phi_3} ~=~r \cdot \phi \ ,
~~~~~~~~r \in \mathbb{R}\ ,\ee
with the normalization $\sum_i \phi^\dagger_i \phi_i^{} = 1$,
the value of the factor $r$ that minimizes $V$ is given by
\be
r^2 ~=~ \frac{m^2}{2(\lambda + \kappa E)} \ ,
\ee
where 
\be
E ~ \equiv ~ \left\{\begin{array}{ll}
\mathrm{Max}\,(\phi_1^3 \phi_3^{} + \phi_3^3 \phi_2^{} + \phi_2^3 \phi_1^{} ~+~
\mathrm{c.c.}) &  \kappa < 0 \ , \\[3mm]  
\mathrm{Min}\, (\phi_1^3 \phi_3^{} + \phi_3^3 \phi_2^{} +\phi_2^3 \phi_1^{} ~+~
\mathrm{c.c.}) &  \kappa > 0 \ . \end{array}\right. 
\ee

\noindent Let us first discuss the case $\kappa < 0$. We are looking for a
maximum of Klein's invariant with fixed normalization for the triplet
$\phi$. Adopting the ansatz 
\be
\pmatrix{\phi_1 \cr \phi_2 \cr \phi_3} ~=~ \frac{1}{\sqrt{3}} \cdot
\pmatrix{e^{i \theta_1} \cr e^{i \theta_2}  \cr e^{i \theta_3}} \ ,
\ee
we readily find
$$ (\phi_1^3 \phi_3^{} + \phi_3^3 \phi_2^{} + \phi_2^3 \phi_1^{} ~+~
\mathrm{c.c.}) ~=~ \frac{2}{9} \Big[ \cos{(3\theta_1} + \theta_3)
~+~\cos{(3\theta_3} + \theta_2) ~+~\cos{(3\theta_2} + \theta_1) \Big] \ .
$$
This takes its maximum if, with $n_i \in \mathbb{N} $,
\begin{eqnarray}
3\theta_1 + \theta_3  =  2 \pi n_1 \ ,~~~~~
3\theta_2 + \theta_1  =  2 \pi n_2 \ ,~~~~~
3\theta_3 + \theta_2  =  2 \pi n_3 \ ,~~~~~
\end{eqnarray}
which we can rewrite as
\begin{eqnarray*}
\theta_1 & = & \frac{\pi}{14} \, (n_2 - 3 n_3 + 9 n_1) \ , \\
\theta_2 & = & \frac{\pi}{14} \, (n_3 - 3 n_1 + 9 n_2) \ , \\
\theta_3 & = & \frac{\pi}{14} \, (n_1 - 3 n_2 + 9 n_3) \ . 
\end{eqnarray*}
We now reparametrize this solution. We start with $\theta_1$ which,
due to the parameter $n_2$, can take any multiple of ${\pi}/{14}$. Let us
therefore define the new integer parameter
\be
t ~\equiv ~ n_2 - 3 n_3 +9 n_1 \ ,
\ee
and replace $n_2$ in the above equations by 
$$
n_2 ~=~  t + 3 n_3 - 9 n_1 \ .
$$
We obtain
\begin{eqnarray}
\theta_1 & = & \phantom{-3\,}t\cdot\frac{\pi}{14}  \ , \nonumber \\
\theta_2 & = & \phantom{-}9\,t\cdot\frac{\pi}{14}  ~-~6\pi n_1 ~+~ 2 \pi n_3\ , \nonumber\\
\theta_3 & = & -3\,t \cdot\frac{\pi}{14} ~+~ 2\pi n_1 \ . \nonumber
\end{eqnarray}
As $2\pi$ rotations of the angles $\theta_i$ are meaningless, the most
general solution for our ansatz is
\be
(\theta_1 \,,\,\theta_2 \,,\,\theta_3) ~=~ \frac{\pi}{14} \, (t \,,\, 9t
\,,\, -3t) \ , ~~~~~t=0,...,27 \ .
\ee
These are 28 different solutions.
The case with $\kappa >0$  can be easily traced back to this result
by observing that Klein's quartic changes its sign under the phase
transformation $\phi_i \,\rightarrow \, e^{{i \pi}/{4}} \, \phi_i \ .$

\noindent Defining the phase factor
\be
\epsilon ~\equiv e^{{i\pi}/{14}} \ ,
\ee
the 28 vacuum configurations can be rewritten as

\be
\left\{ \frac{1}{\sqrt{3}}\cdot\! \pmatrix{\epsilon^t  
\cr \epsilon^{9t} \cr\epsilon^{-3t}} \Bigg| \,t=0,...,27 \right\} = \:    
\frac{i^k}{\sqrt{3}} \left\{ \pmatrix{1 \cr 1  \cr 1 } ,
\pmatrix{\epsilon^{\pm 1}  \cr \epsilon^{\pm 9}\cr\epsilon^{\mp 3}} ,
\pmatrix{\epsilon^{\mp 3}  \cr \epsilon^{\pm 1}\cr\epsilon^{\pm 9}} ,
\pmatrix{\epsilon^{\pm 9} \cr \epsilon^{\mp 3} \cr\epsilon^{\pm 1}} 
\right\} , 
\ee

\vspace{2mm}

\noindent with $k=0,...,3$. Acting with the group generators on these vectors, we obtain
new alignments. A tedious calculation yields the following set 

\be
~i^k \left\{\!
\pmatrix{1 \cr 1 \cr 1}\!,\!
\pmatrix{i\epsilon^{\pm 1}\cr i\epsilon^{\pm 9} \cr i\epsilon^{\mp 3} }\!,\!
\pmatrix{i a_1\cr i a_2\cr i a_3}\!,\!
\pmatrix{a_1\epsilon^{\pm 1}\cr a_2\epsilon^{\pm 9}\cr a_3\epsilon^{\mp 3}}\!,\!
\pmatrix{a_1\epsilon^{\mp 3}\cr a_2\epsilon^{\pm 1}\cr a_3\epsilon^{\pm 9}}\!,\!
\pmatrix{a_1\epsilon^{\pm 9}\cr a_2\epsilon^{\mp3}\cr a_3\epsilon^{\pm 1}} \!,
\; \mathrm{cycl.\:perm.} \right\}, 
\ee

\vspace{2mm}

\noindent where normalization factors have been neglected and 
\be a_n ~\equiv ~1~-~\cos{\left(\frac{2 n \pi}{7}\right)} \ .\ee
Thus we end up with $4\!\cdot\! (1 + 3 \cdot 9  ) = 112$ different vacuum
alignments that minimize the potential.

The question arises whether the degeneracy of these solutions is
lifted if we go to higher order invariants. We therefore insert the above
vacuum configurations for $\kappa <0$ into the invariants  and find
\begin{eqnarray}
(I^{\,[6]}_{} +\bar{I}^{\,[6]}_{}) \Big|_\mathrm{vac} &=&
\frac{4}{27} \cdot \left\{\begin{array}{ll} -1 & ~~ k ~ \mathrm{even}\ , \\ +1
    & ~~ k ~ \mathrm{odd} \ , \end{array} \right.  \\
(I^{\,[14]}_{} +\bar{I}^{\,[14]}_{} )\Big|_\mathrm{vac} &=&
\frac{32}{729}\cdot \left\{\begin{array}{ll} -1 & ~~ k ~ \mathrm{even}\ , \\ +1
    & ~~ k ~ \mathrm{odd} \ , \end{array} \right.   \\
(I^{\,[21]}_{} +\bar{I}^{\,[21]}_{})\Big|_\mathrm{vac}&=&0\ .
\end{eqnarray}
So with $\kappa < 0$, the 112-fold degeneracy transforms into a 56-fold one.
On the other hand, with $\kappa > 0$ we get 
\be
(I^{\,[6]}_{} +\bar{I}^{\,[6]}_{} )\Big|_\mathrm{vac}
=~ (I^{\,[14]}_{} +\bar{I}^{\,[14]}_{})\Big|_\mathrm{vac}
=~ (I^{\,[21]}_{} +\bar{I}^{\,[21]}_{}) \Big|_\mathrm{vac} = ~0 \ .
\ee
Therefore, the 112-fold degeneracy persists in this case 
when including the higher order invariants.


\vskip .4cm
\noindent{\bf Other Quartic Invariants}
\vskip .2cm

\noindent  As an illustration, we now consider quartic invariants made out of
two triplets and two sextets.  They are easy to form, starting from the cubic invariants. We have
\be
 \left[L^{\,i \alpha}_{~~\,a}\,\,   \varphi^{}_i\, \chi^{}_\alpha\, \right]\m D^{ab}_{} \left[L^{\,j\beta}_{~~\,b}\,\,   \varphi'_j\, \chi'_\beta\, \right]\ ,\ee
as well as
\be
 \left[M^{\,i\,\alpha}_{~~\,A} \,\,   \varphi^{}_i\, \chi^{}_\alpha\, \right]\m \delta_{AB}^{} \left[M^{\,j\,\beta}_{~~\,\,B} \,\,   \varphi'_j\, \chi'_\beta\, \right]\ ,\ee
both of which contain the same fields. An explicit calculation shows that both
invariants are independent of each other and totally symmetric. We have

\bea
\!\!&\!\!\frac{1}{3}\!\!&\!\! \Big(
\varphi^{}_{1} \, \varphi'_{1} \; \chi^{}_{6} \, \chi'_{6} ~+~
\varphi^{}_{2} \, \varphi'_{2} \; \chi^{}_{5} \, \chi'_{5} ~+~
\varphi^{}_{3} \, \varphi'_{3} \; \chi^{}_{4} \, \chi'_{4} \Big)
\nonumber \\
+\!\!&\!\!\frac{1}{3}\!\!&\!\!\Big(
\varphi^{}_{1} \, \varphi'_{2} \; \chi^{}_{5} \, \chi'_{6} ~+~
\varphi^{}_{1} \, \varphi'_{3} \; \chi^{}_{4} \, \chi'_{6} ~+~
\varphi^{}_{2} \, \varphi'_{3} \; \chi^{}_{4} \, \chi'_{5} ~+~\mathrm{sym.}\Big) 
\nonumber \\
+\!\!&\!\! \frac{\sqrt{2}}{6}\!\!&\!\! \Big(
\varphi^{}_{1} \, \varphi'_{2} \; \chi^{}_{3} \, \chi'_{4} ~+~
\varphi^{}_{1} \, \varphi'_{3} \; \chi^{}_{2} \, \chi'_{5} ~+~
\varphi^{}_{2} \, \varphi'_{3} \; \chi^{}_{1} \, \chi'_{6} ~+~\mathrm{sym.}\Big)
\nonumber \\
+ \!\!&\!\!\frac{1}{6}\!\!&\!\! \Big(
\varphi^{}_{1} \, \varphi'_{1} \; \chi^{}_{2} \, \chi'_{3} ~+~
\varphi^{}_{2} \, \varphi'_{2} \; \chi^{}_{1} \, \chi'_{3} ~+~
\varphi^{}_{3} \, \varphi'_{3} \; \chi^{}_{1} \, \chi'_{2}~+~ \mathrm{sym.}\Big) 
\nonumber \\
-\!\!&\!\!\frac{1}{\sqrt{2}}\!\!&\!\! \Big(
\varphi^{}_{1} \, \varphi'_{1} \; \chi^{}_{1} \, \chi'_{4} ~+~
\varphi^{}_{2} \, \varphi'_{2} \; \chi^{}_{2} \, \chi'_{6} ~+~
\varphi^{}_{3} \, \varphi'_{3} \; \chi^{}_{3} \, \chi'_{5} ~+~\mathrm{sym.}\Big)
\nonumber \\ 
-\!\!&\!\!\frac{1}{2}\!\!&\!\! \Big(
\varphi^{}_{1} \, \varphi'_{2} \; \chi^{}_{1} \, \chi'_{1} ~+~
\varphi^{}_{1} \, \varphi'_{3} \; \chi^{}_{3} \, \chi'_{3} ~+~
\varphi^{}_{2} \, \varphi'_{3} \; \chi^{}_{2} \,\chi'_{2}~+~\mathrm{sym.}\Big)  \ ,
\eea
and
\bea
\!\!&\!\!\frac{2}{3}\!\!&\!\! \Big(
\varphi^{}_{1} \, \varphi'_{1} \; \chi^{}_{6} \, \chi'_{6} ~+~
\varphi^{}_{2} \, \varphi'_{2} \; \chi^{}_{5} \, \chi'_{5} ~+~
\varphi^{}_{3} \, \varphi'_{3} \; \chi^{}_{4} \, \chi'_{4} \Big)
\nonumber \\
-\!\!&\!\!\frac{1}{3}\!\!&\!\!\Big(
\varphi^{}_{1} \, \varphi'_{2} \; \chi^{}_{5} \, \chi'_{6} ~+~
\varphi^{}_{1} \, \varphi'_{3} \; \chi^{}_{4} \, \chi'_{6} ~+~
\varphi^{}_{2} \, \varphi'_{3} \; \chi^{}_{4} \, \chi'_{5} ~+~\mathrm{sym.}\Big) 
\nonumber \\
+\!\!&\!\! \frac{\sqrt{2}}{3}\!\!&\!\! \Big(
\varphi^{}_{1} \, \varphi'_{2} \; \chi^{}_{3} \, \chi'_{4} ~+~
\varphi^{}_{1} \, \varphi'_{3} \; \chi^{}_{2} \, \chi'_{5} ~+~
\varphi^{}_{2} \, \varphi'_{3} \; \chi^{}_{1} \, \chi'_{6} ~+~\mathrm{sym.}\Big)
\nonumber \\
- \!\!&\!\!\frac{2}{3}\!\!&\!\! \Big(
\varphi^{}_{1} \, \varphi'_{1} \; \chi^{}_{2} \, \chi'_{3} ~+~
\varphi^{}_{2} \, \varphi'_{2} \; \chi^{}_{1} \, \chi'_{3} ~+~
\varphi^{}_{3} \, \varphi'_{3} \; \chi^{}_{1} \, \chi'_{2}~+~ \mathrm{sym.}\Big)  \ ,
\eea
where ``sym.'' indicates symmetrization in the triplet and the 
sextet indices (if they are not already symmetric). Explicitly, for $i\neq j$
and $\alpha \neq \beta$,
$$
\varphi^{}_{i} \, \varphi'_{j} \; \chi^{}_{\alpha} \, \chi'_{\beta}~+~
\mathrm{sym.} ~=~ 
\varphi^{}_{i} \, \varphi'_{j} \; \chi^{}_{\alpha} \, \chi'_{\beta}~+~
\varphi^{}_{i} \, \varphi'_{j} \; \chi^{}_{\beta} \, \chi'_{\alpha}~+~
\varphi^{}_{j} \, \varphi'_{i} \; \chi^{}_{\alpha} \, \chi'_{\beta}~+~
\varphi^{}_{j} \, \varphi'_{i} \; \chi^{}_{\beta} \, \chi'_{\alpha} \ ,
$$
and, for $i = j$ and $\alpha \neq \beta$, 
$$
\varphi^{}_{i} \, \varphi'_{i} \; \chi^{}_{\alpha} \, \chi'_{\beta}~+~
\mathrm{sym.} ~=~ 
\varphi^{}_{i} \, \varphi'_{i} \; \chi^{}_{\alpha} \, \chi'_{\beta}~+~
\varphi^{}_{i} \, \varphi'_{i} \; \chi^{}_{\beta} \, \chi'_{\alpha} \ .
$$
We can understand the fact that these two quartic invariants are symmetric in
the triplets and the antitiplets by looking at a rearranged version of
products of representations. Using the Kronecker products, we obtain
\bea
({\bf 3} \otimes {\bf 6}) \otimes ({\bf 3} \otimes {\bf 6}) 
&=& ({\bf 3} \otimes {\bf 3}) \otimes ({\bf 6} \otimes {\bf 6}) \nonumber \\
&=& \Big({\bf \overline3}_a + {\bf 6}_s\Big) \otimes \Big( 
({\bf 1} + {\bf 6} +{\bf 6} +{\bf 8})_s + ({\bf 7} + {\bf 8} )_a \Big)  \ .\nonumber
\eea
From this we can construct invariants only in two ways, namely
through the symmetric sextets. 

Another example is the quartic invariant 
\be \left[\left(\lambda^{}_{A}\right)^{\,i}_{~\, j} \varphi^{}_i \,\overline\varphi^{j}_{}\right]\delta^{}_{AB}\,\left[\left(\lambda^{}_{B}\right)^{\,k}_{~\, l} \varphi'_k \,{\overline\varphi'}^l\right]\ .\ee
Using the identity
\be
\left(\lambda^{}_{A}\right)^{\,i}_{~\,j}\,\left(\lambda^{}_{A}\right)^{\,k}_{~\, l}~=~2\,\delta^i_{~\, l}\delta^k_{~\, j}-\frac{2}{3}\,\delta^i_{~\, j}\delta^k_{~\, l}\ ,\ee
this invariant reduces to a product of quadratic invariants.

The following table shows the number of independent quartic invariants which
include at least two triplets (for the chiral fermions?), i.e. which are of
type ${\bf 3} \otimes {\bf 3} \otimes {\bf r} \otimes {\bf s}$.
\begin{center}
\begin{tabular}{c|c|c|c|c|c} 
$\phantom{\Big|}({\bf r} , {\bf s})~$ & $~{\bf 3}~$ &  $~{\bf \overline 3}~$ & $~{\bf 6}~$ & $~{\bf  7}~$ & $~{\bf 8}~$ \\ \hline
$\phantom{\Big|}{\bf 3}~$ & 1 & 0 & 0 & 1 & 2 \\ \hline
$\phantom{\Big|}{\bf \overline 3}~$ && 2 & 1 & 1 & 1 \\ \hline
$\phantom{\Big|}{\bf 6}~$ &&& 2 & 2 & 3 \\ \hline
$\phantom{\Big|}{\bf 7}~$ &&&& 3 & 3 \\ \hline
$\phantom{\Big|}{\bf 8}~$ &&&&& 3
\end{tabular}
\end{center}

\vspace{1cm}

\section{The Fano Plane Representation}
We have seen how  $\mathcal{PSL}_2(7)$ acts on Klein's quartic curve through its three-dimensional representation. Its six-dimensional representation also has a remarkable geometrical meaning: it  acts on  the Fano projective plane, the simplest finite projective plane with seven points and seven lines. 
 
In order to see this, consider seven points, imaginatively labeled as $1,2,\dots,7$.  
We wish to construct the group formed by those permutations that permute the seven points {\em and} the seven columns 

$$\begin{array}{cccccccc}
&\left[\begin{array}{c} 1\\ 2\\ 4 \end{array}\right]
&\left[\begin{array}{c} 2\\ 3\\ 5 \end{array}\right]
&\left[\begin{array}{c} 3\\ 4\\ 6 \end{array}\right]
&\left[\begin{array}{c} 4\\ 5\\ 7 \end{array}\right]
&\left[\begin{array}{c} 5\\ 6\\ 1 \end{array}\right]
&\left[\begin{array}{c} 6\\ 7\\ 2 \end{array}\right]
&\left[\begin{array}{c} 7\\ 1\\ 3 \end{array}\right]\\
\end{array}$$ 
A convenient description of these arrangements is the Fano Plane where each triplet of points forms a line

\vskip .3cm
\setlength{\unitlength}{.35in}
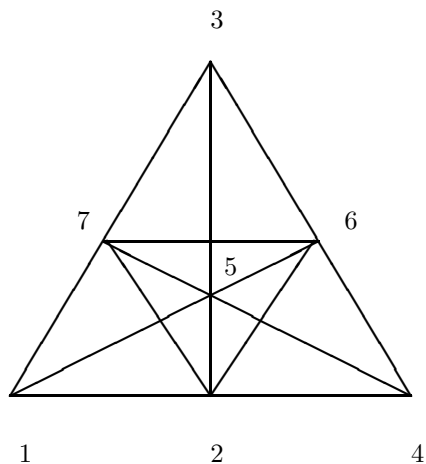
\begin{picture}(8,8)
\thicklines
\put(4,1){\line(1,0){6}}
\put(4,1){\line(2,1){4.6}}
\put(4,1){\line(3,5){3}}
\put(10,1){\line(-3,5){3}}
\put(10,1){\line(-2,1){4.6}}
\put(7,1){\line(0,1){5}}
\put(7,1){\line(2,3){1.55}}
\put(7,1){\line(-2,3){1.55}}
\put(4,0){ 1}
\put(7,0){2}
\put(10,0){4}
\put(7,6.5){3}
\put(7.2,2.8){5}
\put(5,3.5){7}
\put(9,3.5){6}
\put(5.4,3.31){\line(1,0){3.23}}
\end{picture}
\vskip .3cm

\noindent This picture is often used to identify the octonion structure functions. It  is also useful in identifying the subgroup that leaves the line  $[1\,2\,4]$ invariant. We can permute the remaining points 
without changing the picture: the three transpositions $(6\,7)$, $(5\,7)$, and $(3\,7)$ do not alter the picture but relabel the points while 
keeping the base of the triangle invariant. For example, $(6\,7)$ simply interchanges the two outer edges of the triangle. All three generate 
the permutations on the four points 3, 5, 6, 7:  the subgroup that leaves one line invariant is the permutation group on four objects,  $\m S_4$.    

Now consider a transformation that maps the columns into one another, such as the order-seven permutation
\be
c~=~(1234567)\ ,\qquad c^7=1\ ,\ee
which maps the first column into the second one. Similarly, $c^2$ maps the first column into the third one, and so on. This produces the group we are 
seeking through the coset decomposition
\be
\m G~=~\m H+\m H c+ \cdots + \m H c^6\ .\ee
This yields our group of order $7\cdot 24=168$. Geometrically, we can think of the Fano plane as describing a geometry with seven points and seven lines, such that two points are {\em only} on one line, and two lines cross at only one point. It is the smallest example of a finite projective plane.    

The action of $\mathcal{PSL}_2(7)$ on the Fano plane yields a real representation in terms of permutations 
\bea
&& AB=(1234567)\ ,\quad B=(124)(3)(567)\ ,\nonumber \\[2mm]
&& A=(AB)B^2=(1)(5)(6)(23)(47)\ .
\eea
Since  
 
 {\footnotesize
 \be A_{}^{Fano}=\pmatrix{1&0&0&0&0&0&0\cr
 0&0&1&0&0&0&0\cr
 0&1&0&0&0&0&0\cr
 0&0&0&0&0&0&1\cr
 0&0&0&0&1&0&0\cr
 0&0&0&0&0&1&0\cr
 0&0&0&1&0&0&0} , \qquad 
 B_{}^{Fano}=\pmatrix{0&1&0&0&0&0&0\cr
 0&0&0&1&0&0&0\cr
 0&0&1&0&0&0&0\cr
 1&0&0&0&0&0&0\cr
 0&0&0&0&0&1&0\cr
 0&0&0&0&0&0&1\cr
 0&0&0&0&1&0&0}
\ . \ee
 }

\noindent  These two matrices clearly  satisfy the $\mathcal{PSL}_2(7)$ presentation. We can form its order four element  
 
 {\footnotesize
 \be
 [A,B]^{Fano}_{}=\pmatrix{ 0&0&0&0&1&0&0\cr
 0&0&0&1&0&0&0\cr
 0&1&0&0&0&0&0\cr
 0&0&0&0&0&0&1\cr
 1&0&0&0&0&0&0\cr
 0&0&0&0&0&1&0\cr
0&0&1&0&0&0&0} .
\ee
 }
 Their traces
   
\be 
 \Tr\, A^{Fano}_{}=3\ ,\qquad \Tr\, B^{Fano}_{}=1\ ,\qquad \Tr\,[A,B]^{Fano}_{}=1\ ,\ee

\noindent  do not conform to the characters of the septet:  the Fano plane representation is reducible, with 
$$
{\bf 7}^{Fano}_{}~=~\bf 1~+~\bf 6\ .$$
The singlet corresponds to the overall labeling of the seven points, and has no physical content. The same  occurs in the Tetrahedral group, $\m A_4$ which acts on the four vertices of the tetrahedron, yielding a reducible representation of its triplet and singlet representations. 
\section{Outlook}
Mathematicians have long recognized in the simple group $\mathcal{PSL}_2(7)$ a remarkably rich structure with important applications in
algebra, geometry and number theory.\footnote{Klein's Quartic Curve is immortalized in a sculpture at the entrance of the Mathematical Sciences Institute in Berkeley.} However, its beauty has not yet
penetrated the consciousness of physicists. We hope that our presentation will
serve as a physicists' gateway to its further studies. Its maximal subgroups have already been used more or less successfully to
describe the family structure of quarks and leptons. It is therefore
tempting to study how $\mathcal{PSL}_2(7)$ could be applied as a flavor
group. A first step in this direction has been taken in this article by
determining the extreme values of Klein's quartic invariant, for which potentially CP-violating phases  naturally  appear. Its
geometrical incarnation as a Riemann surface of genus three might also
open new paths for string compactification.

\section*{Acknowledgments}
We are grateful for inspiring discussions with G. G. Ross.
PR acknowledges the hospitality of the Aspen Center for Physics
where part of this work was done. CL is supported by the 
University of Florida through the Institute for Fundamental Theory. SN and
PR are supported by the Department of Energy Grant No. DE-FG02-97ER41029.

\newpage 
\noindent{\Large\bf Appendix A}

\vskip .7cm
\noindent{\Large\bf  Tables of Clebsch-Gordan Coefficients}
\vskip .4cm

\noindent In this appendix, we tabulate the Clebsch-Gordan coefficients of $\m P \m S \m
L_2(7)$ which involve triplets and antitriplets. First, we summarize our
notation

\begin{center}
$
\begin{array}{lllcl}
{\bf 3} \: \otimes \: {\bf 3} ~=~ {\bf 6} \,:& ~~~~ & 
   |\, {\alpha} \, \} &=&  K^{ij}_{~~\,\alpha} ~ |\,i>|\,j> \ ,\\[1mm]
{\bf 3} \: \otimes \: {\bf 6} ~=~ {\bf 7} \,:& ~~~~ & 
   |\,a \succ &=&  L^{i\alpha}_{~~\,a} ~ |\,i>|\,\alpha\,\} \ ,\\[1mm]
{\bf 3} \: \otimes \: {\bf 6} ~=~ {\bf 8} \,:& ~~~~ & 
   |\,A \,) &=&  M^{i\alpha}_{~~\,A} ~ |\,i>|\,\alpha\,\} \ ,\\[1mm]
{\bf 3} \: \otimes \: {\bf 7} ~=~ {\bf 7} \,:& ~~~~ & 
   |\,a \succ  &=&  N^{ib}_{~~\,a} ~ |\,i>|\,b\succ \ ,\\[1mm]
{\bf 3} \: \otimes \: {\bf 7} ~=~ {\bf 8} \,:& ~~~~ & 
   |\,A \,)  &=&  P^{ia}_{~~\,A} ~ |\,i>|\,a\succ \ ,\\[1mm]
{\bf 3} \: \otimes \: {\bf 8} ~=~ {\bf 8} \,:& ~~~~ & 
   |\,A \,)  &=&  Q^{i}_{\,BA} ~ |\,i>|\,B\,) \ .\\[1mm]
\end{array}
$
\end{center}
The numerical values for $K^{ij}_{~~\,\alpha}$ can be found in
Section~4.2, while those for $M^{i\alpha}_{~~\,A}$,
$L^{i\alpha}_{~~\,a}$, $N^{ib}_{~~\,a}$, $P^{ia}_{~~\,A}$, $Q^{i}_{\,BA}$ are
listed in the following tables. From Table~A-1 we obtain, for example,
$$
|\,8\,) ~=~ M^{i\alpha}_{~~\,8} |\,i>|\,\alpha\,\} ~=~ \frac{1}{\sqrt{2}}\,
|\,2>|\,5\,\}  ~-~ \frac{1}{\sqrt{2}}\, |\,1>|\,6\,\} \ .
$$
 
\vspace{1mm}

\noindent Notice that we do not give the Clebsch-Gordan coefficients for 
products like $\bf 3  \otimes  7 = 6\,$ since they are directly related to
the previously derived decomposition $\bf 3  \otimes  6 = 7$.
This can be seen easily by rewriting the invariant  
$$
( L^{i \alpha}_{~~a} \varphi_i \chi_\alpha ) \, \mathcal{D}^{ab} \, \psi_b ~=~ 
(L^{i a}_{~~\alpha}\varphi_i\psi_a )\,\mathcal{C}^{\alpha \beta} \, \chi_\beta
\ ,
$$
where we have defined $L^{i a}_{~~\alpha}  \equiv 
\mathcal{C}_{\alpha \beta}  \mathcal{D}^{ab} L^{i \beta}_{~~b}\,$. 

\vspace{2mm}

\noindent Similarly, the Clebsch-Gordan coefficients for products like 
${\bf{\overline{3} \otimes 6 = 7}}$ can be obtained from $\bf 3  \otimes  6 =
7$. Starting with the invariant
\begin{eqnarray}
(L^{i\alpha}_{~~a} \varphi_i \chi_\alpha)  \psi^a \ ,\nonumber
\end{eqnarray}
its complex conjugate is also an invariant:
\begin{eqnarray}
({L^{i\alpha}_{~~a}}^\ast {\varphi_i}^\ast {\chi_\alpha}^\ast ) 
 {\psi^a}^\ast \ .\nonumber
\end{eqnarray}
Since ${\varphi_i}^\ast$ transforms like a general $\bf{\overline{3}}$ field ${\varphi^\prime}^i$, ${\chi_\alpha}^\ast$ like $\mathcal{C}^{\alpha\beta} {\chi^\prime}_\beta \equiv {\chi^\prime}^\alpha$, and ${\psi^a}^\ast$ like $\mathcal{D}_{ab} {\psi^\prime}^b \equiv {\psi^\prime}_a$, we get the invariant
\begin{eqnarray}
({L^{i\alpha}_{~~a}}^\ast {\varphi^\prime}^i {\chi^\prime}^\alpha)
{\psi^\prime}_a  \ ,  \nonumber
\end{eqnarray}
suggesting the definition 
$${L^{i\alpha}_{~~a}}^\ast \equiv L^{~~a}_{i\alpha}  \ .$$
Thus we find
\begin{eqnarray}
( {L_{i\beta}^{~~b}}  {\varphi^\prime}^i \mathcal{C}^{\alpha\beta}
{\chi^\prime}_\alpha )  \mathcal{D}_{ab}{\psi^\prime}^a   
~=~ ( \mathcal{C}^{\alpha\beta}\mathcal{D}_{ab}  {L_{i\beta}^{~~b}}  
{\varphi^\prime}^i {\chi^\prime}_\alpha )  {\psi^\prime}^a   
~=~ 
 ( L^{\:\alpha}_{i\;a} {\varphi^\prime}^i {\chi^\prime}_\alpha  )  {\psi^\prime}^a \ . \nonumber
\end{eqnarray}

\begin{center}
$
\begin{array}{c||cccccccc}
\multicolumn{1}{c||}{\phantom{\Big|}  {\bf{3}}\otimes{\bf{6}}~} 
 & \multicolumn{8}{c}{ {\bf{8}} } \\\hline
 \phantom{\Big|}|\: i > \: |\: \alpha \, \} ~ &  |\: 1 \, )  &  |\: 2 \, )  &  |\: 3 \, )  &  |\: 4 \, )  &  |\: 5 \, )  &  |\: 6 \, ) &  |\: 7 \, )  &  |\: 8 \, ) 
\\\hline\hline &&&&&&&&\\[-3mm]
\phantom{|}|\: 1 > \: |\: 1 \, \} ~  & 0 & 0 & 0 & 0 & 0 & 0 & 0 & 0     \\[1mm]  
\phantom{|}|\: 1 > \: |\: 2 \, \} ~  & 0 & 0 & 0 & 0 & 0 & \sqrt{\frac{1}{3}} & -\sqrt{\frac{1}{3}}i & 0     \\[1mm]  
\phantom{|}|\: 1 > \: |\: 3 \, \} ~  & 0 & 0 & 0 & 0 & 0 & -\sqrt{\frac{1}{3}} & -\sqrt{\frac{1}{3}}i & 0     \\[1mm]  
\phantom{|}|\: 1 > \: |\: 4 \, \} ~  & 0 & 0 & 0 & \sqrt{\frac{1}{6}} & -\sqrt{\frac{1}{6}}i & 0 & 0 & 0     \\[1mm]  
\phantom{|}|\: 1 > \: |\: 5 \, \} ~  & -\sqrt{\frac{1}{6}} & \sqrt{\frac{1}{6}}i & 0 & 0 & 0 & 0 & 0 & 0     \\[1mm]  
\phantom{|}|\: 1 > \: |\: 6 \, \} ~  & 0 & 0 & \sqrt{\frac{1}{6}} & 0 & 0 & 0 & 0 & -\sqrt{\frac{1}{2}}     \\[1mm]  
\phantom{|}|\: 2 > \: |\: 1 \, \} ~  & 0 & 0 & 0 & -\sqrt{\frac{1}{3}} & \sqrt{\frac{1}{3}}i & 0 & 0 & 0     \\[1mm]  
\phantom{|}|\: 2 > \: |\: 2 \, \} ~  & 0 & 0 & 0 & 0 & 0 & 0 & 0 & 0     \\[1mm]  
\phantom{|}|\: 2 > \: |\: 3 \, \} ~  & 0 & 0 & 0 & \sqrt{\frac{1}{3}} & \sqrt{\frac{1}{3}}i & 0 & 0 & 0     \\[1mm]  
\phantom{|}|\: 2 > \: |\: 4 \, \} ~  & 0 & 0 & 0 & 0 & 0 & -\sqrt{\frac{1}{6}} & \sqrt{\frac{1}{6}}i & 0     \\[1mm]  
\phantom{|}|\: 2 > \: |\: 5 \, \} ~  & 0 & 0 & \sqrt{\frac{1}{6}} & 0 & 0 & 0 & 0 & \sqrt{\frac{1}{2}}     \\[1mm]  
\phantom{|}|\: 2 > \: |\: 6 \, \} ~  & \sqrt{\frac{1}{6}} & \sqrt{\frac{1}{6}}i & 0 & 0 & 0 & 0 & 0 & 0     \\[1mm]  
\phantom{|}|\: 3 > \: |\: 1 \, \} ~  & \sqrt{\frac{1}{3}} & -\sqrt{\frac{1}{3}}i & 0 & 0 & 0 & 0 & 0 & 0     \\[1mm]  
\phantom{|}|\: 3 > \: |\: 2 \, \} ~ & -\sqrt{\frac{1}{3}} & -\sqrt{\frac{1}{3}}i & 0 & 0 & 0 & 0 & 0 & 0     \\[1mm]  
\phantom{|}|\: 3 > \: |\: 3 \, \} ~  & 0 & 0 & 0 & 0 & 0 & 0 & 0 & 0     \\[1mm]  
\phantom{|}|\: 3 > \: |\: 4 \, \} ~  & 0 & 0 & -\sqrt{\frac{2}{3}} & 0 & 0 & 0 & 0 & 0     \\[1mm]  
\phantom{|}|\: 3 > \: |\: 5 \, \} ~  & 0 & 0 & 0 & 0 & 0 & \sqrt{\frac{1}{6}} & \sqrt{\frac{1}{6}}i & 0     \\[1mm]  
\phantom{|}|\: 3 > \: |\: 6 \, \} ~ & 0 & 0 & 0 & -\sqrt{\frac{1}{6}} & -\sqrt{\frac{1}{6}}i & 0 & 0 & 0    
\end{array}
$
\end{center}
\vskip 0.2cm

\noindent ~~Table~A-1. The Clebsch-Gordan coefficients $M^{i\alpha}_{~~A}$ for
$\,{\bf 3} \:\otimes\: {\bf 6}~=~{\bf 8}$.

\newpage

\begin{center}
$
\begin{array}{c||ccccccc}
\multicolumn{1}{c||}{\phantom{\Big|}  {\bf{3}}\otimes{\bf{6}}~} 
& \multicolumn{7}{c}{ {\bf{7}} }\\\hline
 \phantom{\Big|}|\: i > \: |\: \alpha \, \} ~ &  |\: 1 \succ &  |\: 2 \succ & |\: 3 \succ &  |\: 4 \succ &  |\: 5 \succ &  |\: 6 \succ &  |\: 7 \succ 
\\\hline\hline &&&&&&&\\[-3mm]
\phantom{|}|\: 1 > \: |\: 1 \, \} ~ & 0 & 0 & 0 & -\sqrt{\frac{3}{4}} & 0 & 0 & 0     \\[1mm]  
\phantom{|}|\: 1 > \: |\: 2 \, \} ~ & 0 & 0 & 0 & 0 & 0 & \sqrt{\frac{1}{12}} & 0     \\[1mm]  
\phantom{|}|\: 1 > \: |\: 3 \, \} ~ & 0 & 0 & \sqrt{\frac{1}{3}} & 0 & 0 & 0 & 0     \\[1mm]  
\phantom{|}|\: 1 > \: |\: 4 \, \} ~ & 0 & 0 & 0 & 0 & \sqrt{\frac{2}{3}} & 0 & 0      \\[1mm]  
\phantom{|}|\: 1 > \: |\: 5 \, \} ~ & 0 & 0 & 0 & 0 & 0 & 0 & \sqrt{\frac{1}{6}}    \\[1mm]  
\phantom{|}|\: 1 > \: |\: 6 \, \} ~ & -\sqrt{\frac{1}{3}} & 0 & 0 & 0 & 0 & 0 & 0      \\[1mm]  
\phantom{|}|\: 2 > \: |\: 1 \, \} ~ & 0 & 0 & 0 & 0 & \sqrt{\frac{1}{3}} & 0 & 0     \\[1mm]  
\phantom{|}|\: 2 > \: |\: 2 \, \} ~ & 0 & 0 & 0 & 0 & 0 & 0 & -\sqrt{\frac{3}{4}}     \\[1mm]  
\phantom{|}|\: 2 > \: |\: 3 \, \} ~ & 0 & 0 & 0 & \sqrt{\frac{1}{12}} & 0 & 0 & 0    \\[1mm]  
\phantom{|}|\: 2 > \: |\: 4 \, \} ~ & 0 & 0 & 0 & 0 & 0 & \sqrt{\frac{1}{6}} & 0     \\[1mm]  
\phantom{|}|\: 2 > \: |\: 5 \, \} ~ & -\sqrt{\frac{1}{3}} & 0 & 0 & 0 & 0 & 0 & 0     \\[1mm]  
\phantom{|}|\: 2 > \: |\: 6 \, \} ~ & 0 & \sqrt{\frac{2}{3}} & 0 & 0 & 0 & 0 & 0    \\[1mm]  
\phantom{|}|\: 3 > \: |\: 1 \, \} ~ & 0 & 0 & 0 & 0 & 0 & 0 & \sqrt{\frac{1}{12}}     \\[1mm]  
\phantom{|}|\: 3 > \: |\: 2 \, \} ~ & 0 & \sqrt{\frac{1}{3}} & 0 & 0 & 0 & 0 & 0    \\[1mm]  
\phantom{|}|\: 3 > \: |\: 3 \, \} ~ & 0 & 0 & 0 & 0 & 0 & -\sqrt{\frac{3}{4}} & 0     \\[1mm]  
\phantom{|}|\: 3 > \: |\: 4 \, \} ~ & -\sqrt{\frac{1}{3}} & 0 & 0 & 0 & 0 & 0 & 0    \\[1mm]  
\phantom{|}|\: 3 > \: |\: 5 \, \} ~ & 0 & 0 & \sqrt{\frac{2}{3}} & 0 & 0 & 0 & 0     \\[1mm]  
\phantom{|}|\: 3 > \: |\: 6 \, \} ~ & 0 & 0 & 0 & \sqrt{\frac{1}{6}} & 0 & 0 & 0\end{array}
$
\end{center}
\vskip 0.2cm

\noindent ~~~~~~~~~~Table~A-2. The Clebsch-Gordan coefficients
$L^{i\alpha}_{~~a}$ for $\,{\bf 3} \:\otimes\: {\bf 6}~=~{\bf 7}$.

\newpage

\begin{center}
$
\begin{array}{c||ccccccc}
\multicolumn{1}{c||}{ \phantom{\Big|}  {\bf{3}}\otimes{\bf{7}}~ } 
& \multicolumn{7}{c}{ {\bf{7}} }  \\\hline
 \phantom{\Big|}|\: i > \: |\: a \succ ~  &  |\: 1 \succ  &  |\: 2 \succ  &  |\: 3 \succ  &  |\: 4 \succ  &  |\: 5 \succ  &  |\: 6 \succ  &  |\: 7 \succ  
\\\hline\hline &&&&&&&\\[-3mm]
 \phantom{|}|\: 1 > \: |\: 1 \succ ~ &  0  & {{ -\sqrt{\frac{1}{3}} }}
&  0  &  0  &  0  &  0  &  0  \\
 \phantom{|}|\: 1 > \: |\: 2 \succ ~ &  0  &  0  
& {{ -\sqrt{\frac{{2}}{{3}}} }} &  0  &  0  &  0  &  0  \\
 \phantom{|}|\: 1 > \: |\: 3 \succ ~ &  0  &  0  &  0  
& {{ -\sqrt{\frac{{1}}{{6}}} }} &  0  &  0  &  0  \\
 \phantom{|}|\: 1 > \: |\: 4 \succ ~ &  0  &  0  
&  0  &  0  &  0  &  0  &  0  \\
 \phantom{|}|\: 1 > \: |\: 5 \succ ~ &  0  &  0  
&  0  &  0  &  0  & {{ \sqrt{\frac{1}{6}} }} &  0  \\
 \phantom{|}|\: 1 > \: |\: 6 \succ ~ &  0  &  0  
&  0  &  0  &  0  &  0  & {{ \sqrt{\frac{{2}}{{3}}} }} \\
 \phantom{|}|\: 1 > \: |\: 7 \succ ~ & {{ \sqrt{\frac{1}{3}} }} 
&  0  &  0  &  0  &  0  &  0  &  0  \\
 \phantom{|}|\: 2 > \: |\: 1 \succ ~ &  0  &  0  
& {{ -\sqrt{\frac{1}{3}} }} &  0  &  0  &  0  &  0  \\
 \phantom{|}|\: 2 > \: |\: 2 \succ ~ &  0  &  0  &  0  
& {{ \sqrt{\frac{1}{6}} }} &  0  &  0  &  0  \\
 \phantom{|}|\: 2 > \: |\: 3 \succ ~ &  0  &  0  &  0  
&  0  & {{ -\sqrt{\frac{2}{3}} }} &  0  &  0  \\
 \phantom{|}|\: 2 > \: |\: 4 \succ ~ &  0  &  0  
&  0  &  0  &  0  & {{ \sqrt{\frac{2}{3}} }} &  0  \\
 \phantom{|}|\: 2 > \: |\: 5 \succ ~ &  0  &  0  
&  0  &  0  &  0  &  0  & {{ -\sqrt{\frac{1}{6}} }} \\
 \phantom{|}|\: 2 > \: |\: 6 \succ ~ & {{ \sqrt{\frac{1}{3}} }} 
&  0  &  0  &  0  &  0  &  0  &  0  \\
 \phantom{|}|\: 2 > \: |\: 7 \succ ~ &  0  &  0  
&  0  &  0  &  0  &  0  &  0  \\
 \phantom{|}|\: 3 > \: |\: 1 \succ ~ &  0  &  0  
&  0  &  0  & {{ -\sqrt{\frac{1}{3}} }} &  0  &  0  \\
 \phantom{|}|\: 3 > \: |\: 2 \succ ~ &  0  &  0  
&  0  &  0  &  0  & {{ -\sqrt{\frac{1}{6}} }} &  0  \\
 \phantom{|}|\: 3 > \: |\: 3 \succ ~ &  0  &  0  
&  0  &  0  &  0  &  0  & {{ \sqrt{\frac{1}{6}} }} \\
 \phantom{|}|\: 3 > \: |\: 4 \succ ~ & {{ \sqrt{\frac{1}{3}} }} 
&  0  &  0  &  0  &  0  &  0  &  0  \\
 \phantom{|}|\: 3 > \: |\: 5 \succ ~ &  0  
& {{ -\sqrt{\frac{2}{{3}}} }} &  0  &  0  &  0  &  0  &  0  \\
 \phantom{|}|\: 3 > \: |\: 6 \succ ~ &  0  &  0  
&  0  &  0  &  0  &  0  &  0  \\
 \phantom{|}|\: 3 > \: |\: 7 \succ ~ &  0  &  0  
&  0  & {{ \sqrt{\frac{2}{3}} }} &  0  &  0  &  0  
\end{array}
$
\end{center}
\vskip 0.2cm

\noindent ~~~~~~~~Table~A-3. The Clebsch-Gordan coefficients $N^{i a}_{~~b}$
for $\,{\bf 3} \:\otimes\: {\bf 7}~=~{\bf 7}$.

\newpage

\begin{center}
$
\begin{array}{c||cccccccc}
\multicolumn{1}{c||}{ \phantom{\Big|}  {\bf{3}}\otimes{\bf{7}}~ } 
& \multicolumn{8}{c}{ {\bf{8}} } \\\hline
 \phantom{\Big|}|\: i > \: |\: a \succ ~ 
& \, |\: 1 \: ) \, & \, |\: 2 \: ) \, & \, |\: 3 \: ) \, & \, |\: 4 \: ) \, 
& \, |\: 5 \: ) \, & \, |\: 6 \: ) \, & \, |\: 7 \: ) \, & \, |\: 8 \: ) \, 
\\\hline\hline &&&&&&&&\\[-3mm]
 \phantom{|}|\: 1 > \: |\: 1 \succ ~ 
& {{ \sqrt{\frac{4}{21}} }} & {{ \sqrt{\frac{4}{21}}\,i }} &  0  &  0  &  0  &  0  &  0  &  0  \\
 \phantom{|}|\: 1 > \: |\: 2 \succ ~  
&  0  &  0  &  0  &  0  &  0  & {{ -\sqrt{\frac{2}{21}} }} & {{ -\sqrt{\frac{2}{21}}\,i }} &  0  \\
 \phantom{|}|\: 1 > \: |\: 3 \succ ~
&  0  &  0  &  0  & {{ -\sqrt{\frac{8}{21}} }} & {{ -\sqrt{\frac{8}{21}}\,i }} &  0  &  0  &  0  \\
 \phantom{|}|\: 1 > \: |\: 4 \succ ~ 
&  0  &  0  &  0  & {{ \sqrt{\frac{3}{14}} }} & {{ -\sqrt{\frac{3}{14}}\,i }}  &  0  &  0  &  0  \\
 \phantom{|}|\: 1 > \: |\: 5 \succ ~  
&  0  &  0  &  0  &  0  &  0  & {{ -\sqrt{\frac{2}{21}} }} & {{ \sqrt{\frac{2}{21}}\,i }} &  0 \\
 \phantom{|}|\: 1 > \: |\: 6 \succ ~ 
& {{ -\sqrt{\frac{1}{42}} }} & {{ \sqrt{\frac{1}{42}}\,i }} &  0  &  0  &  0  &  0  &  0  &  0  \\
 \phantom{|}|\: 1 > \: |\: 7 \succ ~ 
&  0  &  0  & {{ -\sqrt{\frac{25}{42}} }} &  0  &  0  &  0  &  0  & {{ \sqrt{\frac{1}{14}} }}\\
 \phantom{|}|\: 2 > \: |\: 1 \succ ~  
&  0  &  0  &  0  &  0  &  0  & {{ \sqrt{\frac{4}{21}} }} & {{ \sqrt{\frac{4}{21}}\,i }} &  0 \\
 \phantom{|}|\: 2 > \: |\: 2 \succ ~ 
&  0  &  0  &  0  & {{ -\sqrt{\frac{2}{21}} }} & {{ -\sqrt{\frac{2}{21}}\,i }} &  0  &  0  &  0  \\
 \phantom{|}|\: 2 > \: |\: 3 \succ ~ 
&  0  &  0  &  0  & {{ -\sqrt{\frac{2}{21}} }} & {{ \sqrt{\frac{2}{21}}\,i }} &  0  &  0  &  0  \\
 \phantom{|}|\: 2 > \: |\: 4 \succ ~ 
&  0  &  0  &  0  &  0  &  0  & {{ -\sqrt{\frac{1}{42}} }} & {{ \sqrt{\frac{1}{42}}\,i }} &  0  \\
 \phantom{|}|\: 2 > \: |\: 5 \succ ~ 
& {{ -\sqrt{\frac{8}{21}} }} & {{ \sqrt{\frac{8}{21}}\,i }} &  0  &  0  &  0  &  0  &  0  &  0  \\
 \phantom{|}|\: 2 > \: |\: 6 \succ ~ 
&  0  &  0  & {{ \sqrt{\frac{1}{42}} }} &  0  &  0  &  0  &  0  & {{ -\sqrt{\frac{9}{14}} }}\\
 \phantom{|}|\: 2 > \: |\: 7 \succ ~
& {{ \sqrt{\frac{3}{14}} }} & {{ \sqrt{\frac{3}{14}}\,i }} &  0  &  0  &  0  &  0  &  0  &  0  \\
 \phantom{|}|\: 3 > \: |\: 1 \succ ~  
&  0  &  0  &  0  & {{ \sqrt{\frac{4}{21}} }} & {{ -\sqrt{\frac{4}{21}}\,i }} &  0  &  0  &  0  \\
 \phantom{|}|\: 3 > \: |\: 2 \succ ~  
&  0  &  0  &  0  &  0  &  0  & {{ -\sqrt{\frac{8}{21}} }} & {{ \sqrt{\frac{8}{21}}\,i }} &  0  \\
 \phantom{|}|\: 3 > \: |\: 3 \succ ~ 
& {{ -\sqrt{\frac{2}{21}} }} & {{ \sqrt{\frac{2}{21}}\,i }} &  0  &  0  &  0  &  0  &  0  &  0  \\
 \phantom{|}|\: 3 > \: |\: 4 \succ ~ 
&  0  &  0  & {{ \sqrt{\frac{8}{21}} }} &  0  &  0  &  0  &  0  & {{ \sqrt{\frac{2}{7}} }}\\
 \phantom{|}|\: 3 > \: |\: 5 \succ ~  
& {{ -\sqrt{\frac{2}{21}} }} & {{ -\sqrt{\frac{2}{21}}\,i }} &  0  &  0  &  0  &  0  &  0  &  0 \\
 \phantom{|}|\: 3 > \: |\: 6 \succ ~
&  0  &  0  &  0  &  0  &  0  & {{ \sqrt{\frac{3}{14}} }}& {{ \sqrt{\frac{3}{14}}\,i }}&  0  \\
 \phantom{|}|\: 3 > \: |\: 7 \succ ~  
&  0  &  0  &  0  & {{ -\sqrt{\frac{1}{42}} }} & {{ -\sqrt{\frac{1}{42}}\,i }} &  0  &  0  &  0  
\end{array}
$
\end{center}
\vskip 0.2cm

\noindent \,~Table~A-4. The Clebsch-Gordan coefficients $P^{i a}_{~~A}$
for $\,{\bf 3} \:\otimes\: {\bf 7}~=~{\bf 8}$.

\newpage

\begin{center}
$
\begin{array}{c||cccccccc}
\multicolumn{1}{c||}{ \phantom{\Big|}  {\bf{3}}\otimes{\bf{8}}~ } 
& \multicolumn{8}{c}{ {\bf{8}} }\\\hline
 \phantom{\Big|}|\: i > \: |\: A \: )   &  |\: 1 \: )  &  |\: 2 \: ) &  |\: 3 \: ) &  |\: 4 \: ) & |\: 5 \: ) &  |\: 6 \: ) &  |\: 7 \: ) &  |\: 8 \: ) 
\\\hline\hline &&&&&&&&\\[-3mm]
\phantom{|}|\: 1 > \: |\: 1 \: ) ~ & 0 & 0 & -\frac{1}{\sqrt{12}} & 0 & 0 & -\frac{1}{\sqrt{12}} & -\frac{i}{\sqrt{12}} & -\frac{1}{2}  \\[3mm]  
\phantom{|}|\: 1 > \: |\: 2 \: ) ~ & 0 & 0 & -\frac{i}{\sqrt{12}} & 0 & 0 & \frac{i}{\sqrt{12}} & -\frac{1}{\sqrt{12}} &  -\frac{i}{2}  \\[3mm]  
\phantom{|}|\: 1 > \: |\: 3 \: ) ~ & \frac{1}{\sqrt{12}} & \frac{i}{\sqrt{12}} & 0 & 0 & 0 & 0 & 0 & 0  \\[3mm]  
\phantom{|}|\: 1 > \: |\: 4 \: ) ~ & 0 & 0 & 0 & 0 & 0 & -\frac{1}{\sqrt{12}} & \frac{i}{\sqrt{12}} & 0  \\[3mm]  
\phantom{|}|\: 1 > \: |\: 5 \: ) ~ & 0 & 0 & 0 & 0 & 0 & -\frac{i}{\sqrt{12}} & -\frac{1}{\sqrt{12}} & 0  \\[3mm]  
\phantom{|}|\: 1 > \: |\: 6 \: ) ~ & \frac{1}{\sqrt{12}} & -\frac{i}{\sqrt{12}} & 0 & \frac{1}{\sqrt{12}} & \frac{i}{\sqrt{12}} & 0 & 0 & 0  \\[3mm]  
\phantom{|}|\: 1 > \: |\: 7 \: ) ~ & \frac{i}{\sqrt{12}} & \frac{1}{\sqrt{12}} & 0 & -\frac{i}{\sqrt{12}} & \frac{1}{\sqrt{12}} & 0 & 0 & 0  \\[3mm]  
\phantom{|}|\: 1 > \: |\: 8 \: ) ~ & \frac{1}{2} & \frac{i}{2} & 0 & 0 & 0 & 0 & 0 & 0  \\[3mm]  
\phantom{|}|\: 2 > \: |\: 1 \: ) ~ & 0 & 0 & 0 & -\frac{1}{\sqrt{12}} & -\frac{i}{\sqrt{12}} & 0 & 0 & 0  \\[3mm]  
\phantom{|}|\: 2 > \: |\: 2 \: ) ~ & 0 & 0 & 0 & \frac{i}{\sqrt{12}} & -\frac{1}{\sqrt{12}} & 0 & 0 & 0  \\[3mm]  
\phantom{|}|\: 2 > \: |\: 3 \: ) ~ & 0 & 0 & 0 & 0 & 0 & -\frac{1}{\sqrt{3}} & -\frac{i}{\sqrt{3}} & 0  \\[3mm]  
\phantom{|}|\: 2 > \: |\: 4 \: ) ~ & \frac{1}{\sqrt{12}} & -\frac{i}{\sqrt{12}} & 0 & 0 & 0 & \frac{1}{\sqrt{12}} & -\frac{i}{\sqrt{12}} & 0  \\[3mm]  
\phantom{|}|\: 2 > \: |\: 5 \: ) ~ & \frac{i}{\sqrt{12}} & \frac{1}{\sqrt{12}} & 0 & 0 & 0 & -\frac{i}{\sqrt{12}} & -\frac{1}{\sqrt{12}} & 0  \\[3mm]  
\phantom{|}|\: 2 > \: |\: 6 \: ) ~ & 0 & 0 & \frac{1}{\sqrt{3}} & -\frac{1}{\sqrt{12}} & \frac{i}{\sqrt{12}} & 0 & 0 & 0  \\[3mm]  
\phantom{|}|\: 2 > \: |\: 7 \: ) ~ & 0 & 0 & \frac{i}{\sqrt{3}} & \frac{i}{\sqrt{12}} & \frac{1}{\sqrt{12}} & 0 & 0 & 0  \\[3mm]  
\phantom{|}|\: 2 > \: |\: 8 \: ) ~ & 0 & 0 & 0 & 0 & 0 & 0 & 0 & 0  \\[3mm]  
\phantom{|}|\: 3 > \: |\: 1 \: ) ~ & 0 & 0 & 0 & \frac{1}{\sqrt{12}} & \frac{i}{\sqrt{12}} & \frac{1}{\sqrt{12}} & -\frac{i}{\sqrt{12}} & 0  \\[3mm]  
\phantom{|}|\: 3 > \: |\: 2 \: ) ~ & 0 & 0 & 0 & \frac{i}{\sqrt{12}} & -\frac{1}{\sqrt{12}} & -\frac{i}{\sqrt{12}} & -\frac{1}{\sqrt{12}} & 0  \\[3mm]  
\phantom{|}|\: 3 > \: |\: 3 \: ) ~ & 0 & 0 & 0 & \frac{1}{\sqrt{12}}& -\frac{i}{\sqrt{12}} & 0 & 0 & 0  \\[3mm]  
\phantom{|}|\: 3 > \: |\: 4 \: ) ~ & -\frac{1}{\sqrt{12}} & -\frac{i}{\sqrt{12}} & -\frac{1}{\sqrt{12}} & 0 & 0 & 0 & 0 &  \frac{1}{2}  \\[3mm]  
\phantom{|}|\: 3 > \: |\: 5 \: ) ~ & -\frac{i}{\sqrt{12}} & \frac{1}{\sqrt{12}} & \frac{i}{\sqrt{12}} & 0 & 0 & 0 & 0 &  -\frac{i}{2}  \\[3mm]  
\phantom{|}|\: 3 > \: |\: 6 \: ) ~ & -\frac{1}{\sqrt{12}} & \frac{i}{\sqrt{12}} & 0 & 0 & 0 & 0 & 0 & 0  \\[3mm]  
\phantom{|}|\: 3 > \: |\: 7 \: ) ~ & \frac{i}{\sqrt{12}} & \frac{1}{\sqrt{12}} & 0 & 0 & 0 & 0 & 0 & 0  \\[3mm]  
\phantom{|}|\: 3 > \: |\: 8 \: ) ~ & 0 & 0 & 0 & -\frac{1}{2} & \frac{i}{2} & 0 & 0 & 0 
\end{array}
$
\end{center}
\vskip 0.1cm

\noindent ~~~~Table~A-5. The Clebsch-Gordan coefficients $Q^{i}_{~AB}$
for $\,{\bf 3} \:\otimes\: {\bf 8}~=~{\bf 8}$.

\newpage



\begin{thebibliography}{99}


\bibitem{su3}
  T.~L. Curtright and P.~G.~O. Freund, in {\em Supergravity}, edited by P. van
     Nieuwenhuizen and D. Z. Freedman (North-Holland, Amsterdam, 1979); 
  M.~Gell-Mann, P.~Ramond, and R.~Slansky, {\em ibid.}; 
  M.~J. Bowick and P.~Ramond, {\em Phys. Lett.}, B103:338, 1981.

\bibitem{su2} F.~G\"ursey and G.~Feinberg, {\em Phys. Rev.}, 128:378, 1962;
  T.~D.~Lee, {\em Nuovo Cim.},  35:975, 1965;
  S.~Meshkov and S.~P.~Rosen, {\em Phys. Rev. Lett.},  29:1764, 1972.

\bibitem{S3}
  S.~Pakvasa and H.~Sugawara, {\em Phys. Lett.}, B73:61, 1978; 
  J.~Kubo, A.~Mondragon, M.~Mondragon and E.~Rodriguez-Jauregui, {\em Prog.
    Theor. Phys.}, 109:795, 2003, Erratum: {\em ibid.} 114:287, 2005,
    {arXiv:hep-ph/0302196};
  F.~Caravaglios and S.~Morisi, 2005, {hep-ph/0503234};
  W.~Grimus and L.~Lavoura, {\em JHEP}, 0508:013, 2005, {hep-ph/0504153};
  W.~Grimus and L.~Lavoura, {\em JHEP}, 0601:018, 2006, {hep-ph/0509239};
  R.~N. Mohapatra, S.~Nasri, and H.-B. Yu, {\em Phys. Lett.}, B639:318, 2006,
    {hep-ph/0605020};  
  R.~Jora, S.~Nasri, and J.~Schechter, {\em Int. J. Mod. Phys.}, A21:5875,
    2006, {hep-ph/0605069};
  O.~Felix, A.~Mondragon, M.~Mondragon and E.~Peinado, {\em Rev. Mex. Fis.}
     S52:67, 2006, {arXiv:hep-ph/0610061};
  Y.~Koide, {\em Eur. Phys. J.}, C50:809, 2007, {hep-ph/0612058};
  K.~S.~Babu, A.~G.~Bachri and Z.~Tavartkiladze, 2007, {arXiv:0705.4419}.



\bibitem{A4}
  E.~Ma and G.~Rajasekaran, {\em Phys. Rev.}, D64:113012, 2001,
    {hep-ph/0106291}; 
  K.~S. Babu, E.~Ma, and J.~W.~F. Valle, {\em Phys. Lett.}, B552:207, 2003,
    {hep-ph/0206292}; 
  G.~Altarelli and F.~Feruglio, {\em Nucl. Phys.}, B720:64, 2005,
    {hep-ph/0504165};
  K.~S. Babu and X.-G. He, 2005, {hep-ph/0507217};
  A.~Zee, {\em Phys. Lett.}, B630:58, 2005, {hep-ph/0508278};
  G.~Altarelli and F.~Feruglio, {\em Nucl. Phys.}, B741:215, 2006,
    {hep-ph/0512103};
  E.~Ma, {\em Mod. Phys. Lett.}  A21:2931, 2006, {arXiv:hep-ph/0607190};
  S.~F. King and M.~Malinsky, {\em Phys. Lett.}, B645:351, 2007,
    {hep-ph/0610250};
  S.~Morisi, M.~Picariello, and E.~Torrente-Lujan, {\em Phys. Rev.},
    D75:075015, 2007, {hep-ph/0702034};
  F.~Feruglio, C.~Hagedorn, Y.~Lin, and L.~Merlo,  {\em Nucl. Phys.},
    B775:120, 2007, {hep-ph/0702194}; 
  M.~Hirsch, A.~S. Joshipura, S.~Kaneko, and J.~W.~F. Valle, 
    {\em Phys. Rev. Lett.}, 99:151802, 2007, {hep-ph/0703046};
  M.~C.~Chen and K.~T.~Mahanthappa, {\em Phys. Lett.}  B652:34, 2007,
    {arXiv:0705.0714};
  A.~Aranda, 2007, {arXiv:0707.3661}.


\bibitem{S4}
  C.~Hagedorn, M.~Lindner, and R.~N. Mohapatra, {\em JHEP}, 06:042, 2006,
   {hep-ph/0602244};
  Y.~Koide, {\em JHEP}, 0708:086, 2007, {arXiv:0705.2275}.

\bibitem{delta27}
  I.~de~Medeiros~Varzielas and G.~G. Ross, {\em Nucl. Phys.}, B733:31, 2006,
    {hep-ph/0507176};
  I.~de~Medeiros~Varzielas, S.~F. King, and G.~G. Ross, {\em Phys. Lett.},
    B644:153, 2007, {hep-ph/0512313};
  I.~de~Medeiros~Varzielas, S.~F. King, and G.~G. Ross, {\em Phys. Lett.},
    B648:201, 2007, {hep-ph/0607045};
  E.~Ma, {\em Mod. Phys. Lett.}, A21:1917, 2006, {hep-ph/0607056};
  C.~Luhn, S.~Nasri, and P.~Ramond, {\em J. Math. Phys.}  48:073501, 2007,
    {hep-th/0701188}; 
  E.~Ma, 2007, {arXiv:0709.0507}.

\bibitem{other}
  D.~B. Kaplan and M.~Schmaltz, {\em Phys. Rev.}, D49:3741, 1994,
    {hep-ph/9311281};
  M.~Schmaltz, {\em Phys. Rev.}, D52:1643, 1995, {hep-ph/9411383};
  C.~Hagedorn, M.~Lindner, and F.~Plentinger, {\em Phys. Rev.}, D74:025007,
    2006, {hep-ph/0604265};
  E.~Ma, 2007, {hep-ph/0701016};
  C.~Luhn, S.~Nasri, and P.~Ramond, {\em Phys. Lett.}, B652:27, 2007,
    { arXiv:0706.2341};
  M.~Frigerio and E.~Ma, {\em Phys. Rev.}, D76:096007, 2007, {arXiv:0708.0166}.

\bibitem{overview}
  E.~Ma, 2007, {arXiv:0705.0327};
  G.~Altarelli, 2007, {arXiv:0705.0860};
  C.~S.~Lam, {\em Phys. Lett.}, B656:193, 2007, {arXiv:0708.3665}.

\bibitem{oldbook}
  G.~A. Miller, H.~F. Blichfeldt, and L.~E. Dickson,
    {\em Theory and Applications of Finite Groups}, John Wiley \& Sons,
    New York 1916, and Dover Edition 1961.
 
\bibitem{Hanany:1998sd}
  A.~Hanany and Y.~H.~He, {\em JHEP}, 9902:013, 1999, {hep-th/9811183}.

\bibitem{patera}
   C.~J.~Cummins and J.~Patera, {\em J. Math. Phys.}, 29:1736, 1988.

\bibitem{McKay}
  J.~McKay, {\em Proc. Symp. Pure Math.}, 37:183, 1980;
  J.~McKay, {\em Proc. Amer. Math. Soc.}, 81:153, 1981;
  R.~Steinberg, {\em Pacific J. Math.}, 118:587, 1985;
  R.~Suter, {\em Manuscr. Math.}, 122:1, 2007, {arXiv:math/0503542}.

\bibitem{Fairbairn}
  W.~M.~Fairbairn, T.~Fulton, and W.~H.~Klink, {\em J. Math. Phys.}, 5:1038,
  1964. 

\bibitem{atlas}
  J.~H.~Conway et.~al., {\em Atlas of Finite Groups}, Oxford University Press,
  1985.

\bibitem{klein}
  F.~Klein, {\em Math. Annalen}, 14:428, 1878/79, english tranlation in 
    S.~Levy, {\em The Eightfold Way: The Beauty of Klein's Quartic Curve},
    Cambridge University Press, 2001. 
\end{thebibliography}
\end{document}